\documentclass{article}
\usepackage{graphicx} 
\usepackage{amsmath}
\usepackage{caption}
\usepackage{subcaption}
\usepackage{amsfonts}
\usepackage{graphicx}
\usepackage{tikz}
\usepackage{pgfplots}
\usepackage{longtable}
\usepackage{lscape} 
\usepackage{comment}
\usepackage{amssymb}
\usepackage{natbib}
\usepackage[margin=1.2in]{geometry}
\usepackage{authblk}
\newtheorem{proposition}{Proposition}
\usepackage{subcaption}
\usepackage{hyperref}

\title{Modelling shock propagation and resilience \\in financial temporal networks}

\author[1,2]{Fabrizio Lillo\thanks{fabrizio.lillo@unibo.it}}
\author[2]{Giorgio Rizzini\thanks{Corresponding Author: giorgio.rizzini@sns.it}}

\affil[1]{Dipartimento di Matematica, Universit\`{a} degli Studi di Bologna, Piazza di Porta San Donato 5, Bologna, Italy}
\affil[2]{Scuola Normale Superiore, Piazza dei Cavalieri 7, Pisa, Italy}

\begin{document}

\maketitle
\begin{abstract}
    Modeling how a shock propagates in a temporal network and how the system relaxes back to equilibrium is challenging but important in many applications, such as financial systemic risk. Most studies so far have focused on shocks hitting a link of the network, while often it is the node and its propensity to be connected that are affected by a shock. Using as starting point the configuration model, a specific Exponential Random Graph model, we propose a vector autoregressive (VAR) framework to analytically compute the Impulse Response Function (IRF) of a network metric conditional to a shock on a node. Unlike the standard VAR, the model is a nonlinear function of the shock size and the IRF depends on the state of the network at the shock time. We propose a novel econometric estimation method that combines the Maximum Likelihood Estimation and Kalman filter to estimate the dynamics of the latent parameters and compute the IRF, and we apply the proposed methodology to the dynamical network describing the electronic Market of Interbank Deposit (e-MID).
\end{abstract}

\section{Introduction}


Understanding how a shock spreads across a system and how it relaxes back to equilibrium is an important, yet hard, and still open task in many complex systems. Recently, the theory of complex networks has been widely adopted  in many different fields (see, for example,  \cite{boccaletti2006,estrada2012,mata2020}) to model the interconnections between the different parts of the system. 
Moreover, an additional challenge is brought about by the dynamical nature of the interactions, leading to the representation of the system in terms of temporal networks. In fact, the shock modifies not only the instantaneous structure of the network, but, more importantly, its dynamics and the appearance of links in future times, since the network reorganizes itself in response to the shock. This makes the problem particularly challenging and requiring a careful modeling.

 Despite the fact that the proposed approach proposed in this paper is general, our main application will be on financial networks. Networks have been widely applied to economics and financial systems where many market participants (e.g. investors, banks, firms) interact with each other with possibly different types of interactions. In this field, the interest for shock propagation and system's resilience is intimately related to the assessment of systemic risk. Since the United States subprime, the European sovereign debt crisis in 2008, and the COVID-19 crisis, it became clear how fragile and interconnected the economic and financial systems are, see, among others, \cite{ackermann2008,sanders2008,allen2010,easley2010} and \cite{caccioli2018} for a review on financial networks and systemic risk. The enormous consequences of a financial shock makes crucial analysing how a system reacts in order to minimize  economic, financial, and social damages. An important aspect is unveiling not only the path with which an external shock propagates within the system, but also quantifying the resilience of the system in terms of shock absorption and recovery time (see, for example, \cite{may2008,battiston2016,holme2012}).  Resilience in complex networks has been studied, for example, in \cite{albert2000, gao2016,ferraro2018,artime2024}, although most of the studies have been focused on resilience of static networks, while we are interested here in temporal networks.

The purpose of this paper is to provide a general modeling framework for studying the propagation of a shock in a temporal network. Unlike other approaches (see, for example, \cite{giraitis2016}), we are interested in studying the evolution of the system when a node (or a subset of nodes) is shocked, rather than considering the case when a link is shocked (for example, it disappears). Since, differently from links, node properties and their propensity to create connections are not always visible, we propose to use a model with node-specific quantities. More specifically, we resort to the celebrated configuration or fitness model (see \cite{holland1981,bianconi2001,caldarelli2002,chatterjee2011,yan2018}), a specific instance of the Exponential Random Graph family (\cite{lusher2013,harris2013}), which associates to each node one (or two in the directed network case) latent variable, termed fitness, describing the propensity of the node to create links. A link between two nodes is created with a probability depending on the fitnesses of the two nodes. Since the original model describes static networks, here we extend it by assuming that the fitnesses evolve according to a vector autoregressive model of order one, VAR(1). Recently \cite{mazzarisi2020} introduced a dynamical version of the fitness model, but they assumed an independent dynamics of each latent variable. On the contrary, our VAR approach allows to model the lagged correlations between latent variables, i.e. how the propensity of a node in creating a link at time $t$ is correlated with the propensity of another node in creating a link at a future time. This is fundamental for modeling shock propagation since it allows to flexibly describe these lagged interactions between nodes.

 Leveraging on the vast literature on VAR models (\cite{lutkepohl2010}), we propose a modification of the Impulse Response Function (IRF) for temporal networks. We remind that, loosely speaking, the IRF describes the expected dynamics of a variable conditional to a shock at initial time. However, differently from standard VAR model, we are interested in studying the IRF for a {\it network metric}, a function of the adjacency matrix describing the graph, such as its density, clustering coefficient, diameter, etc.). To the best of our knowledge, our is the first paper proposing an IRF analysis on a network metric. In particular, the aim is at analyzing the variation of a given network metric in a dynamic setting after that a node's fitness has been shocked. Additionally, we analyze how the system relaxes back to the equilibrium state examining also the recovery time. 

 Since the VAR describes the fitnesses, the IRF of a network metric has very different properties with respect to linear models. First, we show that it is a nonlinear function of the shock size, i.e. doubling the shock size does not lead to doubling the effect. Second, the conditional dynamics of the network metric depends on the initial state of the network, while in standard VAR it is independent from the initial and past state of the system. Thus, in networked system, the timing of the shock has important consequences on the shock's propagation and relaxation dynamics.  

 As mentioned, our approach allows to construct IRF scenarios by considering shocks on nodes. As we detail in Section 2.1, latent variables often correlates with node specific quantities and are more directly affected by shocks. For example, in interbank networks fitnesses strongly correlate with bank exposure and in the World Trade Web they correlate with Gross Domestic Product (GDP). Thus, a shock on a fitness corresponds to a shock on these variables, which clearly describe the propensity of the nodes (banks, country) to create links. Second, in a network with $n$ nodes, the number of latent variables is $O(n)$, while the number of links is $O(n^2)$. Thus a model based on node variables is much more parsimonious than one based on links and presents a much smaller risk of overfitting. Finally, our methodology is flexible enough to describe the situation in which a node at a certain time could delete all its links (as if the node were removed) and at future times it has the chance to connect with other nodes in the network.

 In the last part of the paper, we propose a method to estimate the latent VAR from empirical data. This is needed to construct IRFs for the real problem at hand. 
We propose a novel estimation method of the model by an approach combining maximum likelihood estimation (MLE) method and Kalman filter. Interestingly, a different use of Kalman filter procedure to modelling of Exponential Random Graph parameters' estimations has been recently proposed by \cite{buccheri2023}.
We test the accuracy of our procedure on synthetic networks, confirming the improvement in the accuracy of our methodology in the estimation of latent variables and static parameters. As an empirical illustration,  we estimate the model and we compute the IRF on a dataset describing the evolution of the financial directed network of electronic Market of Interbank Deposit (e-MID), whose topological properties and financial implication have been intensively studied in previous works as \cite{iori2006,iori2008,iori2015,fricke2015a,fricke2015b,barucca2016,temizsoy2017,barucca2018} for its role in systemic risk.


The paper is structured as follows. Section \ref{Methodology} introduces the main background and preliminary concepts of temporal networks and the fitness model as well as the analytical formulation of the expected value of a network metric in a stochastic environment. Section \ref{Section_IRF} presents the general formulation of the Impulse Response Function for a temporal network. Section \ref{IRF_mean_field} analyzes the IRF in a specific solvable framework as well as the role of each model's parameter in shock spreading. Section \ref{empirical_application} introduces the novel econometric procedure based on the Kalman filter to estimate the fitnesses' dynamics, the results of numerical simulations and the application of the model  to the e-MID interbank network.  Conclusions follow. 

\section{Methodology}
\label{Methodology}

{\bf Preliminary concepts on networks.} A directed network is represented by a graph $G=(V,E)$, where $V$ is the set of $n$ nodes and $E \subset V \times V$ is the set of the arcs connecting nodes. We say that two nodes $i,j \in V$ are connected if there is an arc between them, that is if $(i,j) \in E$. We assume that the network is unweighted, that is, if an arc exists, then its weight is one. The topology of the network can be represented by a non-symmetric $n \times n$ matrix adjacency matrix $\mathbf{A}$. The entries of $\mathbf{A}$ are $a_{ij}=1$ if $(i,j)\in E$, zero otherwise. No self-loops are allowed, that is $a_{ii}=0$ for $i = 1,\dots, n$. A network is undirected if $(i,j) \in E$ implies that $(j,i) \in E$ and the adjacency matrix $\mathbf{A}$ is symmetric. A network metric $f(\mathbf{A})$ is a scalar function of the adjacency matrix
$\mathbf{A}$. \\
 As an example, network density measures the proportion of existing links between nodes with respect to the possible ones and is defined as
\begin{equation}
\delta = \frac{\sum_{i,j} a_{ij}}{n(n-1)}.
\label{eq:density}
\end{equation} 
The degree of a node is the number of connections it has with other nodes in the network. In directed graphs, the direction of the arc matters and therefore, the degree of a node can be split into in-degree and out-degree. For any vertex $i$, the in-degree is the number of incoming links while the out-degree is the number of outgoing links. The in- and out-degree can be represented in terms of the adjacency matrix as
$$
d_i^{in}= \mathbf{A}_i^T \vec{\mathbf{1}}, \hspace{4mm} d_i^{out} = \mathbf{A}_i \vec{\mathbf{1}},
$$
where $\mathbf{A}_i$ corresponds to the $i-$th row of the matrix $\mathbf{A}$, $\mathbf{A}^T$ is the transpose of matrix $\mathbf{A}$, and $\vec{\mathbf{1}}$ is a vector of ones of dimension $n$. 
\vspace{2mm}

{\bf Fitness model for static networks.} Fitness (or configuration) model is a latent variable approach to networks. It has been introduced more or less independently many times in the past century, see, among others, \cite{holland1981,bianconi2001,caldarelli2002,chatterjee2011,yan2018}. In static directed networks, each node $i$ is associated with two latent parameters, $\theta^{in}_i$ and $\theta^{out}_i$, termed {\it fitnesses} and describing the propensity of the node to create incoming and outgoing links,  respectively. Then, for each pair of nodes $i$ and $j$, a link from the first to the second is created with probability $g(\theta^{out}_i, \theta^{in}_j)$, where $g$ is the {\it link function}. It is important to stress that arcs are generated independently (even if, obviously, they are not identically distributed). Different functional forms for $g$ can be used. In this paper we use the (standard) logistic form 
\begin{equation}
  g(\theta^{out}_i, \theta^{in}_j):=P(a_{ij} =1| \theta_{i}^{out}, \theta_{j}^{in})= \frac{1}{1+e^{-\theta^{out}_{i}-\theta^{in}_{j}}}, \hspace{4mm} \forall i,j \in V.
 \label{chi_tempo_fissato}
\end{equation}
for directed networks and 
\begin{equation}
  g(\theta_i, \theta_j):=P(a_{ij} =1| \theta_{i}, \theta_{j})= \frac{1}{1+e^{-\theta_{i}-\theta_{j}}}, \hspace{4mm} \forall i,j \in V.
 \label{chi_tempo_fissato_indiretto}
\end{equation}
for undirected networks. With this choice of the link function the domain of the $\theta$s is the whole real axis.
Notice that, when all $\theta$s are the same, the ensemble of networks is the one of the celebrated Erdős-Renyi random graph model. From Equations \eqref{chi_tempo_fissato} and \eqref{chi_tempo_fissato_indiretto}, it is clear that the degree of a node depends strongly on its fitness.

 The choice of the logistic function for $g$ is made because it connects the fitness model with the class of Exponential Random Graphs Models (see \cite{lusher2013,harris2013}). Exponential Random Graphs are probability distributions of graphs belonging to the exponential family, that is
\begin{equation}
P(\mathbf{A}|\boldsymbol{\vec \theta}) = \frac{\exp \big(\sum_{i=1}^q \theta_{i}f_i(\mathbf{A}) \big)}{Z(\boldsymbol{\vec \theta})}.
    \label{Exponential_family_prob}
\end{equation}
$\boldsymbol{\vec \theta} \in {\mathbb R}^q$ is a vector collecting  latent variables which influences the topology, 
 $f_i(\mathbf{A})$ is a network metric which is also a sufficient statistic for the distribution, and $Z(\boldsymbol{\vec \theta})$ is a normalization factor ensuring that $P(\mathbf{A}|\boldsymbol{\vec \theta})$ is a well-defined probability mass function. Choosing the total number of links  (or equivalently the density) as unique sufficient statistic and assuming an undirected and unweighted network, so $q=1$, Eq. \eqref{Exponential_family_prob} reduces to the Erdős-Renyi model, see \cite{renyi1959}. When in an undirected network all degrees of the nodes are considered as a sufficient statistics, i.e.\ $q=n$ and $f_i = d_i$, Eq. \eqref{Exponential_family_prob} reduces to the fitness model for undirected graphs. Finally, when all the in- and out-degrees of the nodes are taken as sufficient statistics, that is $q = 2n$ and $f_i = d_i^{in}$, $f_{i+n} = d_i^{out}$, $i=1,...,n$, Eq. \eqref{Exponential_family_prob} reduces to the directed version of the fitness model of Eq. \eqref{chi_tempo_fissato}.

In the above setting the fitnesses $\theta$s are given.\ In our (temporal) modeling approach below, they are promoted to random variables with a given probability distribution. Thus, the network metrics become random variables depending also on the latent variables $\theta$s and  on their probability distribution $P(\boldsymbol{\vec \theta})$. 
Consider for simplicity an undirected network (the equations easily generalize to the directed case). The expected value of a network metric $f(\mathbf{A})$ is 
\begin{equation}
  {\mathbb E}[f(\mathbf{A})]=\int f(\mathbf{A}) P(\mathbf{A}|\boldsymbol{\vec \theta}) P(\boldsymbol{\vec \theta}) d \boldsymbol{\vec \theta}= \int f(\mathbf{A}) \prod_{i>j}P(a_{ij}| \theta_i, \theta_j) p_{\boldsymbol{\vec \theta}}(\theta_i,\theta_j) d \boldsymbol{\vec \theta} 
  \label{eq:stati_equa_density}
\end{equation}
where $p_{\boldsymbol{\vec \theta}}$ is the bivariate probability density function of $(\theta_i,\theta_j)$. 
Considering as an important example in the following the network density of Eq.\ \eqref{eq:density}, this expression simplifies to 
\begin{equation}
  \mathbb{E}[\delta] = \frac{2}{n(n-1)} \sum_{i > j}\int_{-\infty}^{\infty}\int_{-\infty}^{\infty} \frac{1}{1+e^{-(\theta_i+\theta_j)}} p_{\boldsymbol{\vec \theta}}(\theta_i,\theta_j) d\theta_i d\theta_j.
  \label{expected_density_double_int}
\end{equation}

 In view of the dynamical model we are going to present below, we explicitly consider the case when $p_{\boldsymbol{\vec \theta}}(\theta_i,\theta_j)$ is a bivariate Gaussian where the two variables have mean $m_i$ and $m_j$, respectively, same variance $s^2$, and correlation coefficient $r$. In this case, there is no closed form expression for the double integral in Eq. \eqref{expected_density_double_int}. However, with a change of variables, it can be expressed in terms of the logistic-normal integral 
\begin{equation}
    I(m, s^2) :=  \int_{-\infty}^{\infty} \frac{1}{1+e^{-x}} \frac{1}{s {\sqrt {2\pi }}}e^{-{\frac {1}{2}}({\frac {x-m }{s}})^{2}} dx,
\end{equation}
which is widely used in Bayesian analysis and for which several approximations exist (see \cite{demidenko2013}). It is easy to show that, under the above Gaussian assumptions, it holds 
$$
\int_{-\infty}^{\infty}\int_{-\infty}^{\infty} \frac{1}{1+e^{-(\theta_i+\theta_j)}} p_{\boldsymbol{\vec \theta}}(\theta_i,\theta_j) d\theta_i d\theta_j=I\left(m_i+m_j,2s^2(1+r)\right).
$$
This results can be used to find the average density when all the $\theta$s have the same Gaussian marginal distribution ${\mathcal N}(m,s^2)$ and the correlation coefficient of each pair is $r$:  
\begin{equation}
  \mathbb{E}[\delta] = I(2m, 2s^2(1+r))
  \label{eq:density_homogeneous}
\end{equation}
Calculations are detailed in the Appendix \ref{Appendix_A}.

\begin{figure}
\centering
    \includegraphics[scale = 0.6]{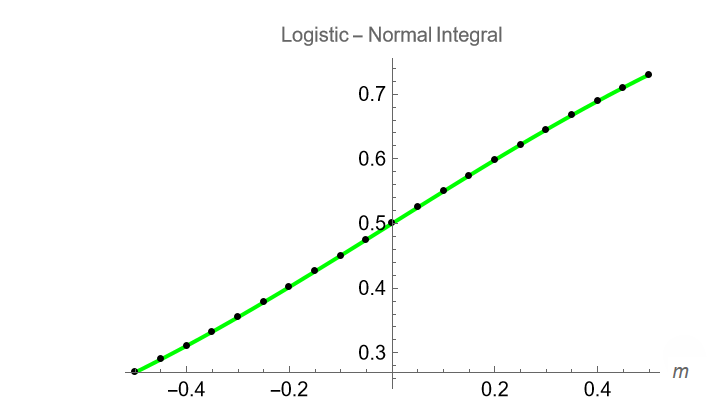}
    \includegraphics[scale = 0.6]{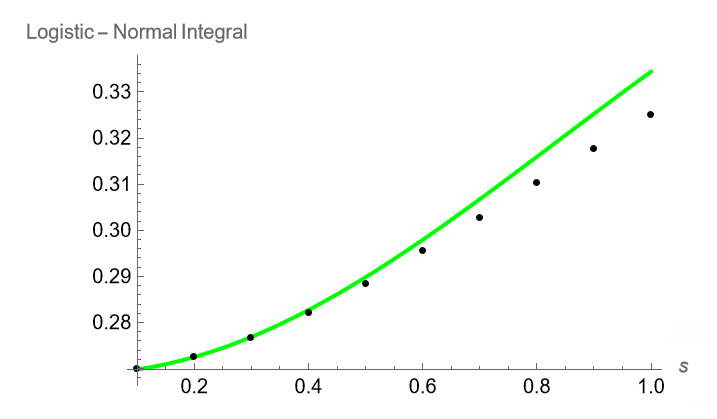}
    \caption{Comparison between exact solution (dotted black line) and second order approximation of the logistic-normal integral (green solid line) in Eq.\ \eqref{logistic_integral}. Parameters: $m = -0.5$, $s = 0.1$, $r = 0$. Left. Comparison with respect to parameter $m$. Right. Comparison with respect to parameter $s$.}
    \label{fig:irf1_confronto}
\end{figure}

 As mentioned above, given its importance in statistics, there are several approximations of the function $I$ when $s$ is small. It is known that the first order approximation performs poorly, so in the following we will use the second order approximation (see \cite{demidenko2013})
\begin{equation}
I(m, s^2) \approx  \frac{1}{1+e^{-m}} \cdot \frac{1}{\sqrt{1+\frac{s^2e^{m}}{(1+e^{m})^2}}} \cdot e^{\frac{s^2}{2((1+e^{m})^2+s^2e^{m})}}.
  \label{logistic_integral}
\end{equation}
Figure \ref{fig:irf1_confronto} shows the comparison between the exact solution of the logistic-normal integral by varying the parameters $m$ and $s$. The plots show that the approximation level is very good for values of $s$ smaller than the absolute mean $|m|$.  

 This result allows to compute the expected density of an undirected network from a fitness model with homogeneous Gaussian distributed $\theta$s, i.e. when$P(\boldsymbol{\vec \theta})$ is a multivariate Gaussian with ${\mathbb E}[\theta_i]=m $, $Var[\theta_i]=s^2$, and $Cor[\theta_i, \theta_j]=r$, $\forall i,j=1,..,n$, as
\begin{equation}
    \mathbb{E}[\delta] \approx \frac{1}{1+e^{-2m}} \cdot \frac{1}{\sqrt{1+\frac{2s^2(1+r)e^{2m}}{(1+e^{2m})^2}}} \cdot e^{\frac{2s^2(1+r)}{2((1+e^{2m})^2+2s^2(1+r)e^{2m})}}.
    \label{density_media_general}
\end{equation}

To shed more light on this result, we perform a Taylor expansion of Eq.\ \eqref{density_media_general} around $s=0$
\begin{equation}
    \mathbb{E}[\delta] \approx \frac{1}{1+e^{-2m}} +(1+\rho) \frac{e^{2m}(1-e^{2m})}{(1+e^{2m})^3}s^2+O(s^4).
\end{equation}  
When $s\to 0$, one obtains the expected density of an Erdős-Renyi model with $p=(1+e^{-2m})^{-1}$. However, for finite $s$, the density depends not only on the expected value of $\theta$ (i.e.\ $m$) but also on its variance and correlation. In particular, when $m<0$ ($m>0$) the expected density increases (decreases) with the variance $s^2$. This effect is amplified by larger correlation coefficients $r$. 

\vspace{2mm}

{\bf Temporal network model.} We now consider a temporal network $\mathcal{G}$ over the time snapshots $\mathcal{T}=\{1,\dots,T\}$ constructed of unweighted and directed networks $G_t$, $t \in \mathcal{T}$. The fitness model can be extended to the temporal networks domain by promoting the the latent variables vector $\boldsymbol{\vec \theta}$ to dynamical variables. Specifically, 
for each time $t\in\mathcal{T}$, each node $i$ is equipped with a couple of latent variables $(\theta_{i,t}^{in}, \theta_{i,t}^{out})$ representing its propensity to create in and out arcs at time $t$, respectively. 
The latent state of the network is described by the vector $\boldsymbol{\vec{{\theta}}}_t \in \mathbb{R}^{2n}$ (or $\mathbb{R}^{n}$ in the undirected case) collecting all the latent variables. Analogously to the static case, these variables are used to generate the network using a link function. The probability that two nodes are linked by an arc at time $t$ is given by a logistic function
\begin{equation}
 g(\theta^{out}_{i,t}, \theta^{in}_{j,t})= P(a_{ij,t} =1| \theta_{i,t}^{out}, \theta_{j,t}^{in})= \frac{1}{1+e^{-\theta^{out}_{i,t}-\theta^{in}_{j,t}}} \hspace{4mm} \forall i,j \in V, \hspace{4mm} \forall t \in \mathcal{T}
 \label{chi_t}
\end{equation}
for a directed network and 
\begin{equation}
 g(\theta_{i,t}, \theta_{j,t})= P(a_{ij,t} =1| \theta_{i,t}, \theta_{j,t})= \frac{1}{1+e^{-\theta_{i,t}-\theta_{j,t}}} \hspace{4mm} \forall i,j \in V, \hspace{4mm} \forall t \in \mathcal{T}
 \label{chi_t_indiretto}
\end{equation}
for an undirected one.

To complete the model it is necessary to specify the dynamics of the vector $\boldsymbol{\vec{{\theta}}}_t$. \cite{mazzarisi2020} proposed a model where each component of $\boldsymbol{\vec{{\theta}}}_t$ evolves independently as an AR(1) process, while \cite{diGangi2022,campajola2022} used a more flexible Score Driven specification to filter a misspecified dynamics, including the possibility of describing non-stationary network dynamics. Importantly, also in this case, each component of $\boldsymbol{\vec{{\theta}}}_t$ evolves independently. 

 Since we are interested in modeling the dependence structure between nodes, in this paper we explicitly consider the case where each latent variable is affected not only by itself but also by other nodes' latent variable. In doing so, we capture the temporal behavior of the latent variables of each node by observing the mutual relationships between them. Specifically, in our model the vector $\boldsymbol{\vec{\theta}}_t$ evolves following a vector autoregressive model of order $1$ (VAR(1)): 
\begin{equation}
    \boldsymbol{\vec{\theta}}_t  = \boldsymbol{\vec{\mu}} +  \mathbf{B}  \boldsymbol{\vec{\theta}}_{t-1} + \boldsymbol{\vec w}_t ,
    \label{Model_VAR_1}
\end{equation}
where, in the undirected case\footnote{For directed network, $\boldsymbol{\vec{\mu}}$  and $\boldsymbol{\vec w}_t$ have $2n$ components while $\mathbf{B}$ is a $2n\times 2n$ matrix.}, $\boldsymbol{\vec{\mu}} \in \mathbb{R}^{n}$ is the vector of constant intercept, $\mathbf{B}$ is a $n\times n$ constant matrix expressing the mutual relations among the latent variables and $\boldsymbol{\vec w}_t \in {\mathbb R}^n$ is the vector of error terms assumed to be Gaussian zero mean white noises. The covariance matrix of $\boldsymbol{\vec w}_t $ is indicated with ${\boldsymbol \Sigma}$. 
As usual in VAR models, covariance stationarity of Eq.\ (\ref{Model_VAR_1}) is guaranteed when the spectral radius (i.e. the maximum of the absolute values of the eigenvalues) of $\mathbf{B}$ is strictly smaller than one.

 The vector $\boldsymbol{\vec{\mu}}$ controls the unconditional mean value of the latent variables, thus the temporal average of the in- and out-degree of each node.  In fact, the steady-state (or, equilibrium state) of the fitness vector is $\boldsymbol{\vec \theta}_S\equiv {\mathbb E}[\boldsymbol{\vec \theta}_t]=(\mathbf{I} - \mathbf{B})^{-1} \boldsymbol{\vec \mu}$ where $\mathbf{I}$ is the identity matrix. The unconditional mean is also controlled by $\mathbf{B}$, which describes the lagged interactions between latent variables. In particular, the diagonal elements of $\mathbf{B}$ describe the temporal autocorrelations of the $\theta$s and thus the autocorrelation of degrees. The off diagonal elements describe how the latent variable of a node at time $t$ affects the latent variable of another node at time $t+1$, thus they are related to the lagged cross correlations between degrees.

\subsection{Impulse Response Analysis and shock propagation}
\label{Section_IRF}

In this section, we propose a modification of the impulse response analysis to study how a shock on a node or on a group of nodes of a temporal network propagates to the other nodes and how the system relaxes back to the equilibrium state. Differently from other approaches which considers a shock on a network observable (e.g.\ a link disappears), we study shocks on nodes' characteristics, namely the fitnesses seen as propensity measures in creating connections. In particular, we investigate the role of the model variables in the determining the severity, breadth, and resilience of a shock. 

The idea of considering the effect of a shock on a latent variable rather than on an observable variable (e.g.\ a link) can be motivated in the following way. First, latent variables often correlates with node specific quantities and are more directly affected by shocks. For example, \cite{mazzarisi2020} shows that  $e^{\theta^{out}_{i,t}}$ and $e^{\theta^{in}_{i,t}}$ of a fitness model estimated in the unweighted interbank network very highly correlate with the contemporaneous bank exposure, i.e. the amount lent and borrowed, respectively, by bank $i$ at time $t$. Thus, a shock $\theta^{out}_{i,t} \to \theta^{out}_{i,t} +\Delta$ corresponds to a log change of $\Delta$ in the amount lent by the bank (or, for small shocks, a $100\cdot  \Delta\%$ change in the exposure). Similarly, in zero-inflated gravity models of the World Trade Web (\cite{burger2009,winkelmann2008})
the probability of a trading link between country $i$ and $j$ is 
$$
P(a_{ij}=1)=\frac{\overline {GDP_i} \cdot \overline{GDP_j}}{1+\overline{GDP_i} \cdot \overline{GDP_j}} 
$$
where $\overline{GDP_i}$ is the (suitably normalized) GDP of country $i$.  Thus we can identify $\theta_i=\log \overline{GDP_i}$ and a shock on $\theta_i$ corresponds to a log change of GDP equal to $\Delta$. Thus a shock of a latent variable (i) is able to model a shock on a node rather than on a link and (ii) captures a possible shock in node intrinsic properties which drive the propensity in creating links.

 A second reason for preferring shocks on nodes rather than on link, can be econometrically motivated by the recent work \cite{diGangi2022}. This paper considers the link prediction problem by comparing a Tobit regression model of a link at time $t$ versus several observable network characteristics at previous times (\cite{giraitis2016}) with a fitness based approach which forecasts the fitnesses of the nodes and from them predicts the link existence. The paper shows that the latter approach outperforms the former, suggesting that to forecast future networks it is better to use a lower dimensional representation of the network (with $n$ fitnesses) rather than a much higher representation in terms of the $n^2$ observable links.

To be specific,  we call $\tau \in \{0,\dots,T\}$ 
the time at which an exogenous shock, $\boldsymbol{\vec{\Delta \theta}}$, occurs in the vector $\boldsymbol{\vec{\theta}}_\tau$. 
We define the (standard) Impulse Response Function (IRF) of the latent vector as
\begin{equation}
{\boldsymbol I}{\boldsymbol R}{\boldsymbol F}^{\boldsymbol{\vec \theta}}(t;\boldsymbol{\vec{\Delta \theta}})= {\mathbb E}[\boldsymbol{\vec{\theta}}_{\tau+t}| \boldsymbol{\vec{\theta}}_{\tau}+\boldsymbol{\vec{\Delta \theta}},\boldsymbol{\vec{\theta}}_{\tau-1}... ]-{\mathbb E}[\boldsymbol{\vec{\theta}}_{\tau+t}| \boldsymbol{\vec{\theta}}_{\tau},\boldsymbol{\vec{\theta}}_{\tau-1}... ]
\label{eq:general_IRF_theta}
\end{equation}
where the expectation is taken over the realizations of the noise $\boldsymbol{\vec  w}_t$. As usual, this represents the expected marginal effect of the shock at time $\tau$ on the future value of the dynamical variable\footnote{Note that in impulse response analysis the shock is often applied to a component of the noise term, while here we shock directly the latent variable.}. Due to the linearity of the VAR model, it is easy to verify (\cite{lutkepohl2005}) that Eq. \eqref{eq:general_IRF_theta} reduces to
\begin{equation}
{\boldsymbol I}{\boldsymbol R}{\boldsymbol F}^{\boldsymbol{\vec  \theta}} (t;\boldsymbol{\vec{\Delta \theta}})=\mathbf{B}^t \boldsymbol{\vec{\Delta \theta}},
\label{eq:irf_theta}
\end{equation}
which is, as expected, a linear function of the shock vector $\boldsymbol{\vec{\Delta\theta}}$. Moreover, as well known in the theory of VAR processes, the IRF does not depend on $\boldsymbol{\vec{\theta}}_{\tau},\boldsymbol{\vec{\theta}}_{\tau-1}...$ but only on the shock vector $\boldsymbol{\vec{\Delta \theta}}$ and on matrix $\mathbf{B}$. 

 However, we are interested in the expected marginal effect of the shock on the properties of the observable network, which is a random function of the vector $\boldsymbol{\vec{\theta}}_t$. Thus, indicating with $f(\mathbf{A}_t)$ a generic network metric (e.g. density, assortativity coefficient, etc) at time $t$, we define the IRF of the metric of the temporal network as 
\begin{equation}
{\boldsymbol I}{\boldsymbol R}{\boldsymbol F}^{f}
(t;\boldsymbol{\vec{\Delta \theta}})={\mathbb E}[f(\mathbf{A}_{t+\tau})| \boldsymbol{\vec{\theta}}_{\tau}+\boldsymbol{\vec{\Delta \theta}},\boldsymbol{\vec{\theta}}_{\tau-1}... ]-{\mathbb E}[f(\mathbf{A}_{t+\tau})| \boldsymbol{\vec{\theta}}_{\tau},\boldsymbol{\vec{\theta}}_{\tau-1}... ]
\label{IRF_general}
\end{equation}
where the average is taken both on the noise of the VAR $\boldsymbol{\vec w}_t$ and on the randomness generating the temporal networks given the latent variables at each time. 

We prove the following:
\begin{proposition} \label{proposition1}
If $\boldsymbol{\vec{\theta}}_t$ follows a VAR(1) dynamics, the Impulse Response Function for the network metric $f(\mathbf{A}_t)$  is
\begin{equation}
{\boldsymbol I}{\boldsymbol R}{\boldsymbol F}^f
(t;\boldsymbol{\vec{\Delta \theta}})=
\int {\mathbb E}\left[f(\mathbf{A}_t)|\boldsymbol{\vec{\theta}}_t\right] 
\cdot \left[ {\mathcal N}\left( \boldsymbol{\vec{\theta}}_t;\boldsymbol{\vec \mu}_t+\mathbf{B}^{t-\tau}\boldsymbol{\vec{\Delta \theta}},\boldsymbol{\Sigma}_t\right)-{\mathcal N}\left( \boldsymbol{\vec{\theta}}_t;\boldsymbol{\vec \mu}_t,\boldsymbol{\Sigma}_t\right)\right]d\boldsymbol{\vec{\theta}}_t
\label{eq:irfp_final}
\end{equation}
where
\vspace{-3mm}
\begin{eqnarray}
\boldsymbol{\vec \mu}_t &=&(\mathbf{I} -\mathbf{B})^{-1}(\mathbf{I} -\mathbf{B})^{t-\tau} \boldsymbol{\vec \mu}  + \mathbf{B}^{t-\tau} \boldsymbol{\vec{\theta}}_\tau \label{eq:mut} \\
\boldsymbol{\Sigma}_t &=&(\mathbf{I} -\mathbf{B}^2)^{-1}(\mathbf{I} -\mathbf{B}^2)^{t-\tau}\boldsymbol{\Sigma}\label{eq:sigmat}
\end{eqnarray}
are, respectively, the conditional mean and variance of $\boldsymbol{\vec \theta}_t|\{\boldsymbol{\vec{\theta}}\}_{0:\tau}$ and $\mathbf{I}$ is the $n\times n$ identity matrix.
\end{proposition}

\medskip 

The proof is presented in Appendix \ref{Appendix_B_new}. Few comments are in order:
\begin{itemize}
\item The expectation ${\mathbb E}\left[f(\mathbf{A}_t)|\boldsymbol{\vec{\theta}}_t\right]$ refers to the {\it static} fitness model. Thus, if one is able to compute the properties of the static model, with a Gaussian integration it is possible to obtain the IRF in response to a shock.
\item Even if $\boldsymbol{\Sigma}$ is diagonal, $\boldsymbol{\Sigma}_t$ is not (unless, of course, also $\boldsymbol{B}$ is diagonal). Thus correlations arise at $t>1$ if the $\theta$s do not evolve independently but influence each other in the VAR. 
\item Differently from the standard linear VAR, the above IRF {\it does} depend on the state of the system at the time of the shock $\boldsymbol{\vec{\theta}}_{\tau}$. Thus, in our model, current history of the network affects its reaction to shocks. 
\item Differently from the IRF on the $\theta$s in Eq.\ \eqref{eq:irf_theta}, the IRF on the network metric is a {\it nonlinear function} of the shock $\boldsymbol{\vec{\Delta \theta}}$. Thus, doubling the shock amplitude generally does not lead to a doubling of the effect on the network metric.
\item Finally, when $t\to \infty$, assuming stationarity of the VAR, it is 
\begin{eqnarray*}
\boldsymbol{\vec \mu}_t &\to& (\mathbf{I}-\boldsymbol{B})^{-1}\boldsymbol{\vec \mu} \\
\boldsymbol{\Sigma}_t &\to& (\mathbf{I}-\boldsymbol{B}^2)^{-1}\boldsymbol{\Sigma} 
\end{eqnarray*}
i.e., as expected, they converge to the stationary value and, therefore, the shock is completely re-absorbed. 
\end{itemize}

 Clearly, it is interesting to investigate how the different model's parameters affect the amplitude of the initial shock and the speed (i.e. the resilience) with which the system returns to the stationary state. Since this in general depends on a very large number of parameters, in the next Section we consider a simple and analytically solvable example.

\section{Impulse Response Function of the mean-field model}
\label{IRF_mean_field}

The dynamics of the VAR model in Section \ref{Methodology} and of the IRF in Section \ref{Section_IRF} strongly depends on the structure of the matrix $\mathbf{B}$. In the following, we  consider different specifications, in the attempt of balancing flexibility with tractability. The matrix $\mathbf{B}$ depends, in fact, on $n^2$ or $4n^2$ parameters for undirected and directed networks, respectively, and some restrictions, both in modelling and in estimation, are in order.

The simplest setting is the one where all the diagonal elements of $\mathbf{B}$ are equal to $a \in \mathbb{R}$ and all the off-diagonal elements are equal to $b\in \mathbb{R}$. When also the components of $\boldsymbol{\vec \mu}$ and the variance of the noise are equal with vanishing covariances, i.e.\ $\boldsymbol{\Sigma} = \sigma^2 \mathbf{I}$, we have a sort of ``mean-field" setting, where all nodes are statistically equivalent and equally affected  by all the other nodes. The largest eigenvalue of the matrix $\mathbf{B}$ is $\lambda_1=a+b(n-1)$ and covariance stationarity of the VAR model is guaranteed when $|\lambda_1|<1$. The stationary value of $\boldsymbol{\vec{\theta}}$ is
\begin{equation}
    \boldsymbol{\vec{\theta}}_S= \frac{\mu}{1-\lambda_1} \boldsymbol{\vec 1}
\end{equation}
i.e. all the $\theta$s have on 
 average the same value.

 A slightly more sophisticated setting assumes that not all the off-diagonal elements of $\mathbf{B}$ are nonvanishing. This describes a situation where the (lagged) interactions are sparse and the dynamics of the latent variable of a node depends only on a subset of other nodes. The model is characterized by an extra parameter $p$ giving the probability with which a non diagonal element of $\mathbf{B}$ is different from zero and equal to $b$. Note that this extra stochasticity is quenched, i.e. the matrix $\mathbf{B}$ is the same at all time steps. Clearly, when $p\to 1$ this model coincides with the mean-field one described above, while for $p\to 0$ only self-interactions are present, reducing the VAR model to a series of $n$ AR(1) processes similarly to \cite{mazzarisi2020}. 

 It is interesting to note that, when, as in the sparse or dense mean-field model, it is $\boldsymbol{\vec{\mu}}=\mu \boldsymbol{\vec{1}}$, then 
\begin{equation}
    \boldsymbol{\vec{\theta}}_S=(\mathbf{I}-\mathbf{B})^{-1}\boldsymbol{\vec \mu}= \mu (\mathbf{I}-\mathbf{B})^{-1}\boldsymbol{\vec 1}
    \label{chain_equality2}
\end{equation}
i.e. the stationary value of the fitness of node is its Katz centrality when the dynamical matrix $\mathbf{B}$, is seen as an adjacency matrix of a weighted network. The Katz centrality (\cite{katz1953}) depends on a tuning parameter $\alpha$, which in Eq.\ \eqref{chain_equality2} is set equal to $1$. Since the largest eigenvalue of $\mathbf{B}$ is smaller than 1, the condition $1=\alpha<\lambda_1^{-1}$ is automatically satisfied for stationary VAR. This interesting result shows that the most central nodes in the interaction network (having $\mathbf{B}$ as weighted adjacency matrix) are the ones with largest degree in the observed stationary network  (having $\mathbf{A}$ as adjacency matrix)

In this section, we derive analytically the IRF in Eq.\ \eqref{eq:irfp_final} for an undirected\footnote{For sake of simplicity and clarity, we focus on an undirected network but, by simple calculations, it is possible to derive also the IRF for a directed network.} network at any time $t$ by considering as a property the network density, i.e. $f(\mathbf{A}_t)=\delta_t$ defined in Eq.\ \eqref{eq:density}. 
Specifically,
we assume to work in the mean-field setting and that the fitness of node $i=1$\footnote{Note that in this homogeneous framework, the choice of the shocked fitness does not influence the IRF.  We select the first node for simplicity.} is shocked at time $\tau =0$ by an amount $\Delta$, i.e. $\boldsymbol{\vec{\Delta \theta}}=(\Delta,0,...,0)'$.
As said above, the ${\boldsymbol I}{\boldsymbol R}{\boldsymbol F}^f(t;\boldsymbol{\vec{\Delta \theta}})$ of Eq.\ \eqref{eq:irfp_final} depends on the state of the system $\boldsymbol{\vec{\theta}}_\tau$ at the time of the shock, $\tau$. To get an intuition on the role of the model's parameters, we consider the case when just before the shock all the $\theta$s have the same value. Of course one could choose it equal to the stationary value $\theta_S$ derived above, but the we will keep it general in order to study how the Impulse Response Function depends on the initial state of the system. 

 The Impulse Response Function in \eqref{eq:irfp_final} depends on the dynamics of the conditional mean and variance of the fitness variables in Equations \eqref{eq:mut} and \eqref{eq:sigmat}, respectively. In  Appendix \ref{Sec_power_matrix_new} we show that the conditional mean and variance can be written in terms of parameters $a$, $b$ and of the spectral radius as
\begin{equation}
    \boldsymbol{\vec \mu}_t = \left(  \frac{1-(a-b)^t}{1-(a-b)} \mathbf{I} +  \left[ \frac{1-\lambda_1^t}{1-\lambda_1} - \frac{1-(a-b)^t}{1-(a-b)} \right]\frac{1}{n} \mathbf{1} \right) \boldsymbol{\vec \mu} +  \left( (a-b)^t \mathbf{I} + \frac{\lambda_1^t - (a-b)^t}{n} \mathbf{1} \right)\boldsymbol{\vec{\theta}}_{0}
    \label{mu_shock_dynamic}
\end{equation}
and 
\begin{equation}
    \boldsymbol{\Sigma}_t = \left(  \frac{1-(a-b)^{2t}}{1-(a-b)^2} \mathbf{I} +   \left[ \frac{1-\lambda_1^{2t}}{1-\lambda_1^2} - \frac{1-(a-b)^{2t}}{1-(a-b)^2} \right]\frac{1}{n} \mathbf{1} \right) \boldsymbol{\Sigma},
    \label{sigma_shock_dynamic}
\end{equation} 
where $\mathbf{1}$ is all-ones matrix, i.e.\ a matrix with elements are one.

As already stated in Section \ref{Methodology}, it is immediate to observe two facts: (i) for $t \ge 2$ the matrix $\boldsymbol{\Sigma}_t$ is no longer diagonal, (ii) the shock intensity does not influence the conditional variance of the fitness variables $\boldsymbol{\Sigma}_t$. Regarding the first point, we label by $\sigma_t = \Sigma_{ii,t}$ and $\rho_t = \Sigma_{ij,t}/\Sigma_{ii,t}$ the conditional variance and the correlation coefficient of each node, respectively. As for the second point, 
the shock intensity $\Delta$ influences the conditional mean of the fitness variables and indicating $\boldsymbol{\vec \mu}^s_t =\boldsymbol{\vec \mu}_t+\mathbf{B}^{t}\boldsymbol{\vec{\Delta \theta}}$, one obtains
\begin{equation}
 \boldsymbol{\vec \mu}^s_t = \left(  \frac{1-(a-b)^t}{1-(a-b)} \mathbf{I} +   \left[ \frac{1-\lambda_1^t}{1-\lambda_1} - \frac{1-(a-b)^t}{1-(a-b)} \right]\frac{1}{n} \mathbf{1} \right)\boldsymbol{\vec \mu} + \left( (a-b)^t \mathbf{I} + \frac{\lambda_1^t - (a-b)^t}{n} \mathbf{1} \right) (\boldsymbol{\vec{\Delta \theta}} + \boldsymbol{\vec{\theta}}_{0})
\label{eq:mu_shock_dinamico}
\end{equation}
The entries of $\boldsymbol{\vec \mu}^s_t$ in \eqref{eq:mu_shock_dinamico} can have different dynamics. Indeed, once the shock is triggered, node $1$ is directly shocked and all the other nodes are shocked indirectly through node $1$.  We label $z$ the generic neighbor of node $1$. We indicate the conditional mean of the shocked node by $\mu^s_{1,t}$, while the conditional mean of all other nodes $z$ by $\mu^s_{z,t}$. They are defined as 
\begin{equation}
\mu^s_{1,t} = \mu    \left( \frac{1-(a-b)^t}{1-(a-b)} +  \left[ \frac{1-\lambda_1^t}{1-\lambda_1} - \frac{1-(a-b)^t}{1-(a-b)} \right]\frac{1}{n} \right) + \theta_0 (a-b)^t + \frac{1}{n} (n\theta_0 + \Delta ) (\lambda_1^t - (a-b)^t) + \Delta  (a-b)^t
\label{eq:mu_i_shocked}
\end{equation}

\begin{equation}
\mu^s_{z,t} = \mu    \left( \frac{1-(a-b)^t}{1-(a-b)} +  \left[ \frac{1-\lambda_1^t}{1-\lambda_1} - \frac{1-(a-b)^t}{1-(a-b)} \right]\frac{1}{n} \right) + \theta_0 (a-b)^t + \frac{1}{n} (n\theta_0 + \Delta) (\lambda_1^t - (a-b)^t).
\label{eq:mu_j_shocked}
\end{equation}
 Eqs.\ \eqref{eq:mu_i_shocked} and \eqref{eq:mu_j_shocked} highlight that the difference between the conditional means is the term $ \Delta  (a-b)^t$ which expresses the impact of a shock $t$ times after  $\tau=0$. In this framework, the Impulse Response Function taking $f(\mathbf{A}_t)= \delta_t$ at any time $t \ge 0$ can be written explicitly in terms of the logistic-normal integral as
\begin{equation}
     {\boldsymbol I}{\boldsymbol R}{\boldsymbol F}^{\delta}(t;\boldsymbol{\vec{\Delta \theta}})= \frac{2}{n}I(\mu_{1,t}^s + \mu_{z,t}^s, 2\sigma_t^2(1+\rho_t))+\frac{n-2}{n}I(2\mu_{z,t}^s,2\sigma_t^2(1+\rho_t))- I(2\mu_t,2\sigma_t^2(1+\rho_t)).
     \label{IRF_temporale}
\end{equation}
The first term describes the effect on the density of shocked node and the other nodes, while the second term describes the indirect effect on density due to the connections between non-shocked nodes. The last term is the baseline density when the shock is not present.

 The immediate effect on the density is described by the above expression when $t=1$. In this case, the IRF can be written as
\begin{equation}
{\boldsymbol I}{\boldsymbol R}{\boldsymbol F}^{\delta}(1;\boldsymbol{\vec{\Delta \theta}})= \frac{2}{n}I(2\mu+2\lambda_1\theta_0+(a+b)\Delta, 2\sigma^2)+\frac{n-2}{n}I(2\mu+ 2\lambda_1\theta_0+2b\Delta_0,2\sigma^2)- I(2\mu+2\lambda_1\theta_0,2\sigma^2).
  \label{eq:IRF_tau_1}
\end{equation}

Finally, when we consider the model with sparse VAR matrix, we will consider averages over different realizations of $\mathbf{B}$. This accounts to replace the off-diagonal elements of $\mathbf{B}$ with $bp$. This is due to the fact that when the elements of the $\mathbf{B}$ are i.i.d. and we indicate with $\mathbb{E}[\mathbf{B}] = \bar{\mathbf{B}}$, then it is direct to show that $\mathbb{E}[\mathbf{B}^k] = \bar{\mathbf{B}}^k$.

\subsection{Numerical analysis} 
\label{numerical_analysis}

To get an intuition on the role of the different model parameters on the relaxation dynamics of the network, we perform in this section a comparative statics using the exact results on the mean-field model presented in the previous section. As a baseline scenario, we consider a network of $n=50$ nodes and we choose $a=0.3$, and $b=0.01$ corresponding to a maximum eigenvalue $\lambda_1 = a+b\cdot (n-1)=0.79$, and we fix the noise variance $\sigma^2=0.1$. Then we consider three values for parameter $\mu$, $\mu \in \{-0.3,0,0.3\}$, corresponding to a ``sparse", ``average" and ``dense" networks, since the corresponding network density is $0.05$, $0.5$ and $0.95$, respectively. We assume that the fitness of all the nodes at the time of the shock are at the stationary value
$\theta_i = \theta_{S}=\mu/(1-\lambda_1)$ $\forall i$, that is $\theta_S \in \{-1.43,0,1.43\}$,  and a shock intensity $\Delta = -10$. A negative value of $\Delta$ corresponds to a sudden drop in the fitness of the shocked node $1$ and as a consequence of the density of the network.
Finally, in all the computations we use the second-order approximated value of $I$ as in Eq.\ \eqref{logistic_integral}. Notice that all the computations in this Section are analytical and follows from the expression derived above.

 \begin{figure}
     \centering    \includegraphics[scale=0.7]{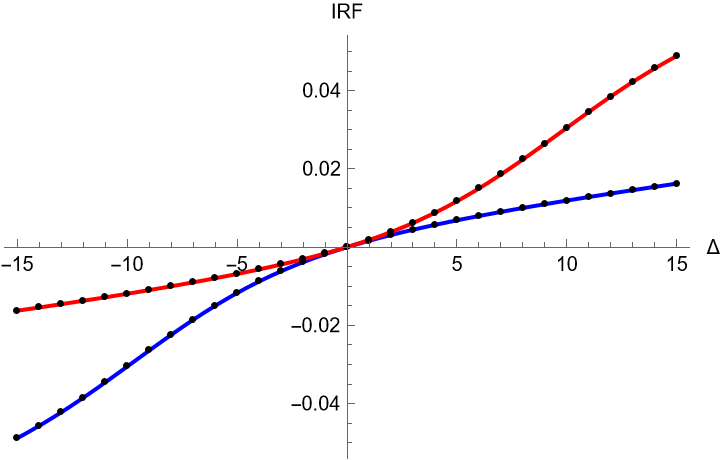}   
     \caption{Dependency of the IRF at time $t=1$ with respect to $\Delta$ with $\mu = -0.3$ (red line) and $\mu = 0.3$ (blue line).}
     \label{fig:irf1}
 \end{figure}

\begin{figure}
    \centering
    \includegraphics[scale = 0.9]{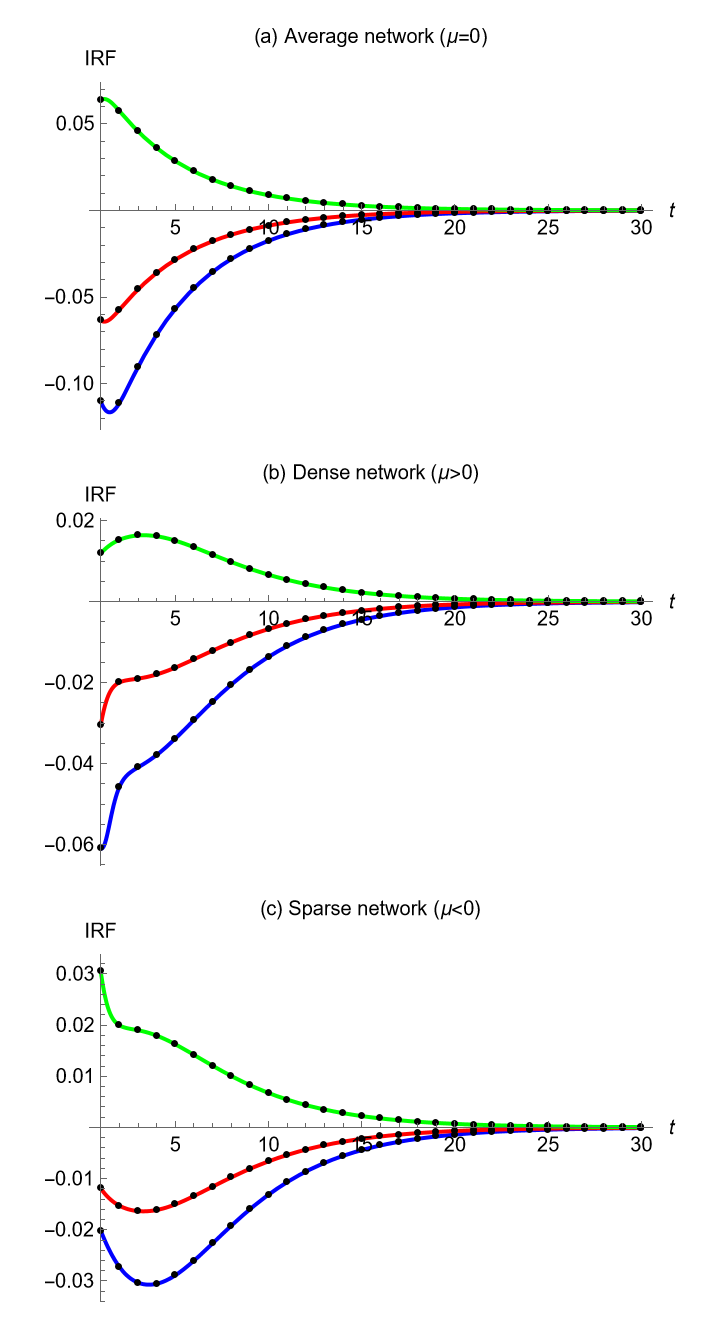}
    \caption{Impulse Response Function with respect to $\Delta$: $\Delta = -20$ (blue line), $\Delta = -10$ (red line) and, $\Delta = 10$ (green line).} \label{fig:IRFt_t_Delta_mu_different}
\end{figure}

\vspace{3mm}
\textbf{Shock intensity $\mathbf{\Delta}$.} 
We first study the role of the shock intensity and therefore, beyond the baseline case $\Delta=-10$, we consider also $\Delta=10$ and $\Delta=-20$. Remind that in the standard VAR, the IRF is proportional to $\Delta$ (see Eq.\ \eqref{eq:irf_theta}), thus it is expected to be symmetric for shocks with the same intensity but opposite sign.
This is indeed observed also for the IRF on the density when $\mu=0$, i.e.\ when the average network density is $0.5$, (see  
panel (a) of Figure \ref{fig:IRFt_t_Delta_mu_different}).
However when $\mu\ne 0$ a clear asymmetric pattern emerges (see Figure \ref{fig:irf1} and panels (b) and (c) of Figure \ref{fig:IRFt_t_Delta_mu_different}). Taking the dense network as an example (panel (b)), a shock of $\Delta =-10$ has an immediate negative effect on the density which is three times larger than a positive shock of $\Delta =10$. When the time elapses, the decay pattern in the two cases becomes more similar in absolute value. A symmetric behavior is observed for sparse networks (panel (c)). This effect can be explained by noticing that when $\mu$ is positive (negative), the network density varies in absolute terms much more with negative (positive) shock than with positive (negative) one. 
This mirroring and asymmetric effect is explained by the presence of the positive (negative) $\mu$, which increases (decreases) the network density with respect to $0.5$. The higher (lower) the parameter $\mu$, the higher (lower) the probability that two nodes are connected and therefore, a positive (negative) will modify the network density negligibly because most of the nodes are already connected (disconnected). On the other hand, when $\Delta$ and $\mu$ have opposite signs, the network density changes significantly. The range of variation depends on $\mu$ and $\Delta$. 
Finally, the non-linear relation between $\mu$ and $\Delta$ causes the system to react to shock differently both in terms of pattern and absorption time. The figure also shows the case $\Delta=-20$ and it is clear that, especially for short times, the pattern of the IRF is different from twice the pattern for $\Delta=-10$, pointing again at a non-linear behavior of the IRF. It is interesting to note that when $\mu$ and $\Delta$ have the same sign, the maximum (minimum) effect might be reached at a time $t > 1$, while in the case of parameters with opposite sign, the IRF is a strictly decreasing (increasing) function. 
We therefore conclude that, in sparse networks, which are frequent in economic and financial applications, the larger disruption of a negative shock is expected to happen at a later time than the shock time.\\
\vspace{3mm}




\begin{figure}
\centering
\includegraphics[scale = 0.9]{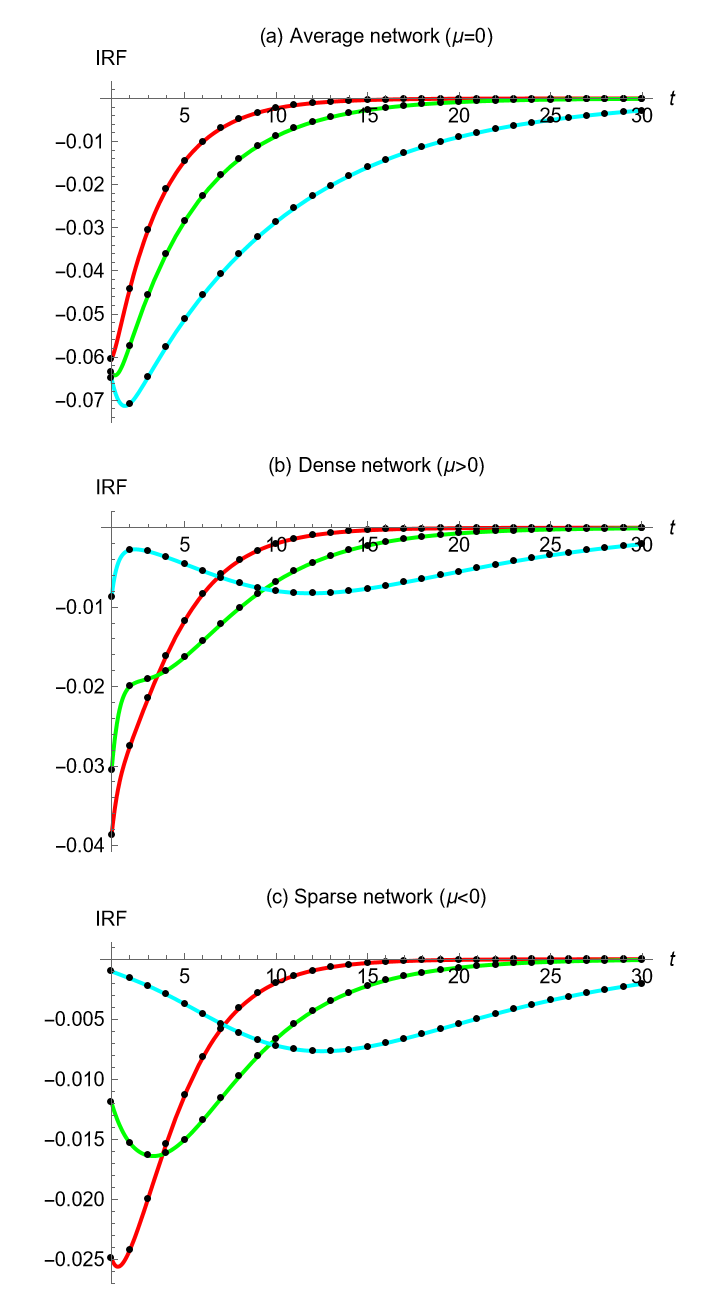}
    \caption{Impulse Response Function with respect to $\lambda_1$: $\lambda_1 = 0.69$ (red line), $\lambda_1 = 0.79$ (green line), and $\lambda_1 = 0.89$ (cyan line). }\label{fig:IRFt_t_a_different_lambda_mu_different}
\end{figure}

\begin{figure}
\centering
\includegraphics[scale = 0.9]{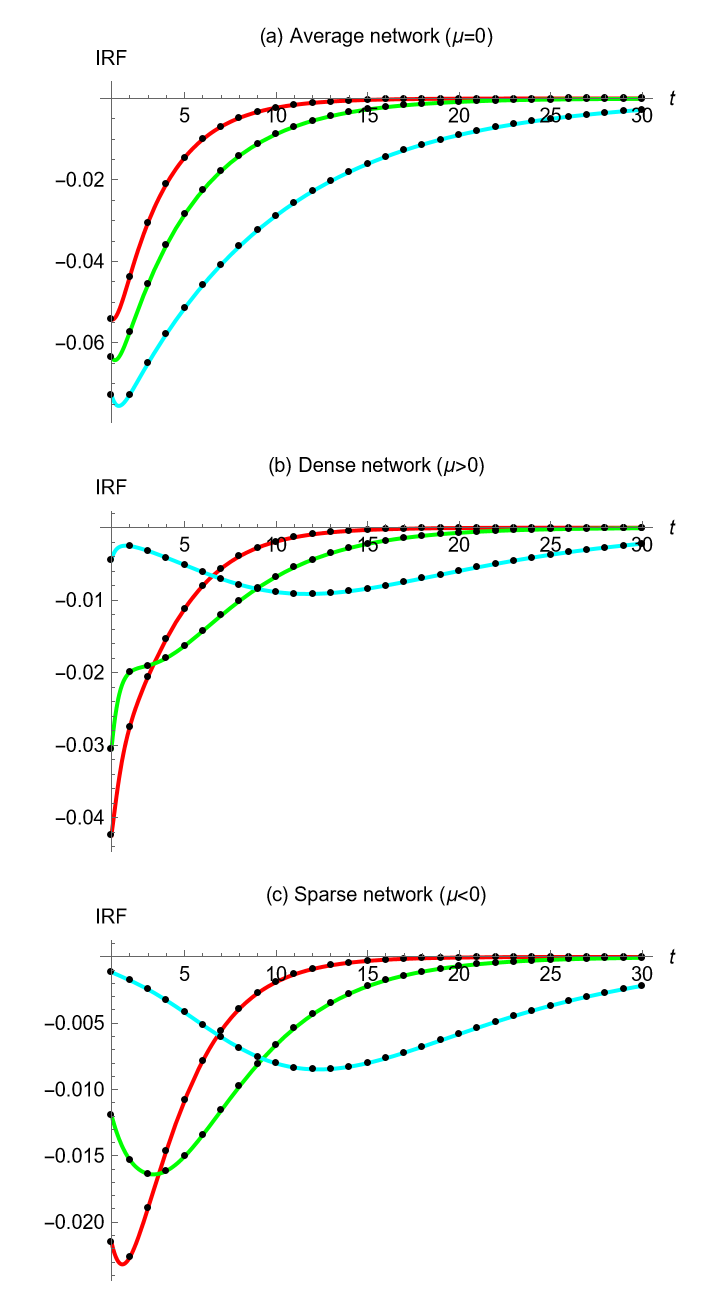}
    \caption{Impulse Response Function with respect to  $b$ keeping $a=0.3$: $b =7.9592\cdot 10^{-3}$ (red line), $b= 0.01$ (green line), and $b=1.2041\cdot 10^{-2}$ (cyan line). }\label{fig:IRFt_t_b_different_lambda_mu_different}
\end{figure}

\begin{figure}
    \centering
    \includegraphics{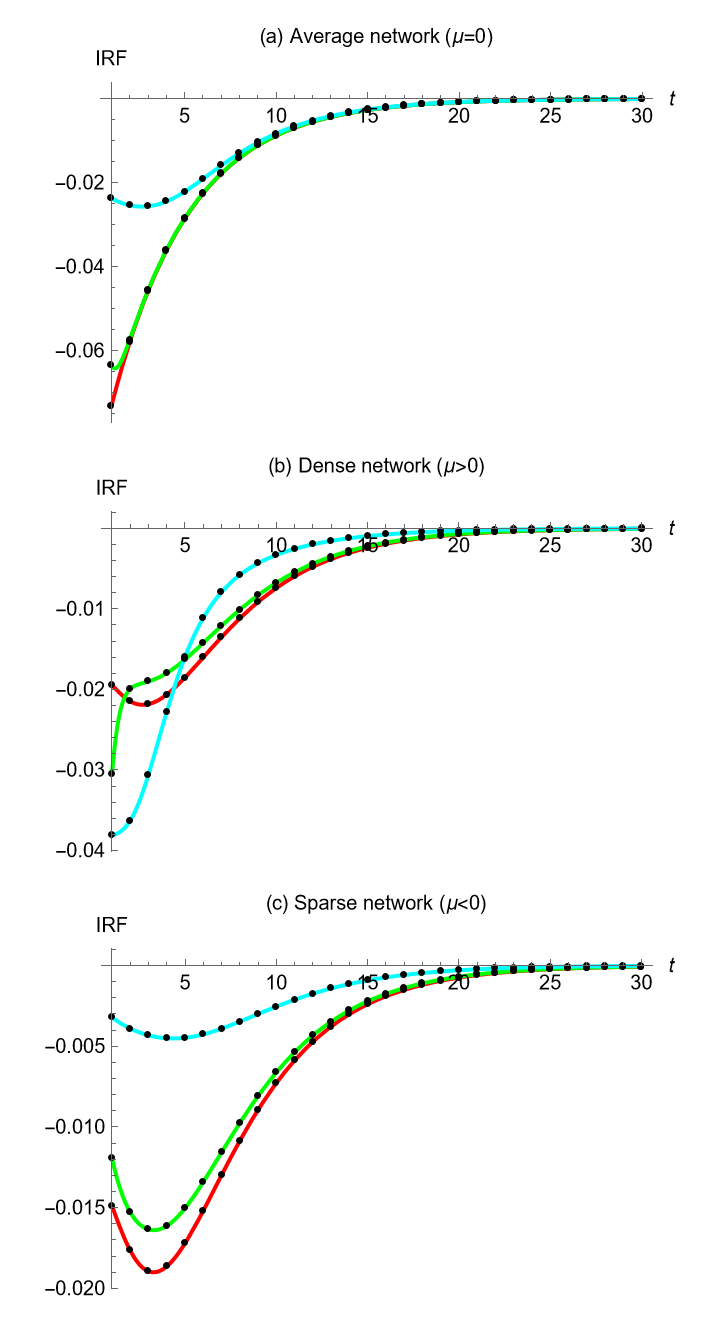}
    \caption{Impulse Response Function with respect to $a$ and $b$ keeping $\lambda_1 = 0.79$: $a=0.05$, $b = 1.5102\cdot 10^{-2}$ (red line), $a = 0.3$, $b=0.01$ (green line) and 
    $a=0.75$, $b= 8.1633\cdot 10^{-4}$ (cyan line).}
    \label{fig:IRFt_t_a_b_same_lambda_mu_different}
\end{figure}

{\bf Spectral radius of $\mathbf{B}$.} Since in the network case the IRF depends in a non-linear way on the powers of matrix $\mathbf{B}$, it is interesting to assess the role of the spectral radius of $\mathbf{B}$ in Eq.\ \eqref{Model_VAR_1} in shock spreading. We modify the spectral radius of matrix $\mathbf{B}$ by changing the parameters $a$ and $b$ such that $\lambda_1 \in \{0.69,0.79,0.89\}$. 
In particular, first we modify the parameter $a$ keeping fixed the parameter $b=0.01$ and later, we vary the parameter $b$ keeping fixed the parameter $a=0.3$. In the first comparative statics, {\color{black}we set} $a\in \{0.2,0.3,0.4\}$ while in the second comparative statics, {\color{black}we choose} $b\in \{7.9592\cdot 10^{-3},0.01,1.2041\cdot 10^{-2}\}$. Results are plotted in Figures \ref{fig:IRFt_t_a_different_lambda_mu_different} and \ref{fig:IRFt_t_b_different_lambda_mu_different} 
both having the baseline as middle case. Based on linear VARs we expect stronger and longer lasting effects of the shock when $\lambda_1$ is larger.
This is indeed observed when $\mu =0$ (panel (a) of Figure \ref{fig:IRFt_t_a_different_lambda_mu_different}). In this case, the results suggest that the higher the spectral radius, the higher is the effect on the network density and the longer is the absorption time\footnote{Indeed, by simple calculations, it is possible to notice that the IRF at time $1$, depends only on the presence of parameter $a$ in the first term of \eqref{eq:IRF_tau_1}.}. More interestingly when $\mu \ne 0$
(panels (b) and (c) of Figure \ref{fig:IRFt_t_a_different_lambda_mu_different}) we observe the opposite effect at short times, i.e. the lower the spectral radius, the larger is the effect of the shock. However the speed of relaxation is faster for smaller spectral radius, as expected. Regarding the shock effect, when $\mu \ne 0$, a change in $\lambda_1$ leads to a different equilibrium value of the fitness. By assuming a positive mean, the higher the value of $a$, the higher the spectral radius and the higher the equilibrium value\footnote{Indeed, for $\mu >0$ and $\lambda_1' < \lambda_1^{''} < \lambda_1^{'''}$, the equilibrium value ordering is
$$
\frac{\mu}{1-\lambda_1^{'}} < \frac{\mu}{1-\lambda_1^{''}} < \frac{\mu}{1-\lambda_1^{'''}}.$$ In case of $\mu <0$, the previous ordering is reversed.}. 
Moreover, for sparse networks the lower the $\lambda_1$, the sooner the IRF attains its maximum effect. This lagged effect can be explained by noticing that, keeping $b$ fixed, increasing $a$, each fitness will be influenced more by itself at the previous time than by the others and therefore the shock requires more time to be transmitted between the fitness and to produce a relevant effect in the network.
 {\color{black}On the contrary}, comparing Figures \ref{fig:IRFt_t_a_different_lambda_mu_different} and \ref{fig:IRFt_t_b_different_lambda_mu_different}, it is possible to notice that the fitnesses' connectedness affects only the IRF in the short-term. This result clearly appear in panel (a) of Figure \ref{fig:IRFt_t_b_different_lambda_mu_different} at $t=1$: the higher the connection between the fitnesses, the higher is the shock effect on the network and therefore, the lower is the IRF. In fact, focusing on panel (a) of Figures \ref{fig:IRFt_t_b_different_lambda_mu_different} and \ref{fig:IRFt_t_a_different_lambda_mu_different}, we observe that the parameter $b$ plays an essential role in determining the IRF at the time immediately after the shock. In fact, in Figure \ref{fig:IRFt_t_a_different_lambda_mu_different}, it is observed that by placing $b=0.01$, the IRFs are concentrated around the value $-0.06$. Using different values of $b$ (and keeping $a$ fixed), the IRFs are in a wider range $(-0.04, 0.07)$.  
In the other panels of Figure \ref{fig:IRFt_t_b_different_lambda_mu_different}, there are no relevant differences in the IRF confirming the predominant role of the parameter $\mu$ in the resilience of the network. \\

{\bf Matrix $\mathbf{B}$ configuration.} 
Although the spectral radius is an important determinant of the IRF's dynamics, it does not contain all the information on the role of $\mathbf{B}$. 
Consequently, we analyse the role of parameters $a$ and $b$ in matrix $\mathbf{B}$ keeping the spectral radius constant to baseline value $\lambda_1 = 0.79$. We consider three cases: (i) high cross-correlation and low autocorrelation, (ii) baseline scenario, (iii) low cross-correlation and high autocorrelation. In the first case, at any time $t$, $\boldsymbol{\vec \theta}_t$ does not depend on its past history but rather on the connections between each fitness. In the third case, $\boldsymbol{\vec \theta}_t$ depends significantly on $\boldsymbol{\vec \theta}_{t-1}$ rather than on the connection between fitness pairs. The case (ii) is an intermediate case. In the first case, the driver of the spectral radius is the parameter $b$, expressing the cross correlation between $\theta$s from $t-1$ to $t$, while in the third case the driver of the spectral radius is the autocorrelation of each fitness variable. The last case can be seen as a proxy of independent fitnesses, i.e.\ the matrix $\mathbf{B}$ becomes a diagonal matrix and the VAR framework reduces to a $n$ independent AR(1) processes. The results are shown in Figure \ref{fig:IRFt_t_a_b_same_lambda_mu_different}. \\
An interesting result is that it is not possible to establish an absolute dependence between the IRF and parameters $a$ and $b$, since the latter depends strongly on the value of $\mu$, which mainly controls the probability of the existence of a link through the link function $g$.  \\
Starting from the case with $\mu = 0$ (panel (a)), it is possible to observe that the higher the connection between couples of $\theta$s, i.e.\ the higher the parameter $b$, the higher is the impact of the shock on the network and, therefore, the lower is the IRF. This result is highlighted also by observing the value of the IRF at time $1$ of cyan curve. 
When $\mu >0$, i.e.\ in presence of a dense network, the previous ordering is reverse: the higher the parameter $a$, the higher is the shock effect on the network and, therefore, the lower is the IRF. This result can be interpreted by observing that in a dense network, the effect on the connection among $\theta$s is negligible since the network is characterized by a high number of links and the presence of additional links (due to the increase of $b$) does not significantly affect the IRF.  On the contrary, when $\mu$ is negative, the highest impact happens when the fitnesses are (almost) fully connected {\color{black}i.e., when $b$ assumes a large value}. Interestingly, it is possible to notice that an increase in parameter $b$ translates to an increasing rate of the connection between the latent variables and, therefore, to the paths' weights through which a shock can be transmitted. Indeed, if we see the matrix $\mathbf{B}$ as the adjacency matrix of a weighted graph, it is easy to observe that the higher is $b$, the higher the link weight and therefore the higher the weight of shorter paths connecting the shocked node and all the other ones. This explains also the reason why in Figure \ref{fig:IRFt_t_a_b_same_lambda_mu_different} the Impulse Response Function reaches its maximum at  $t \ge  1$.
\\

\begin{figure}
\centering
\includegraphics[scale = 0.9]{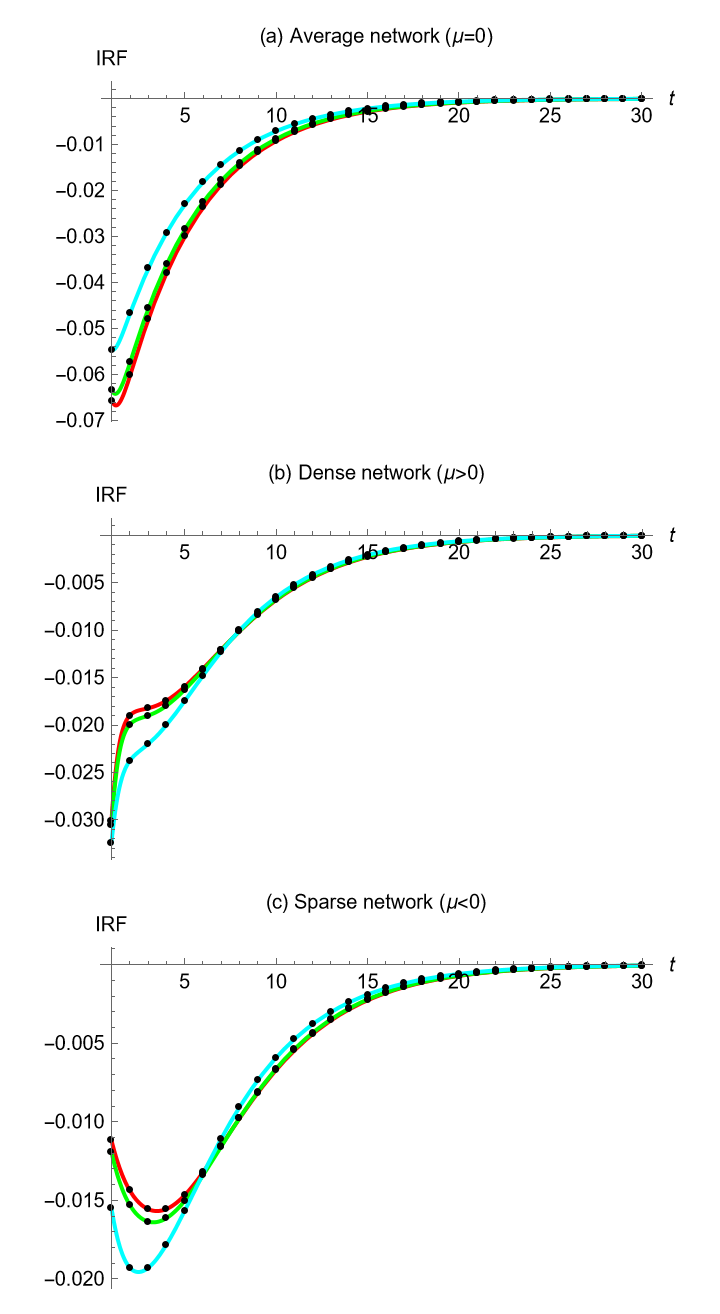}
    \caption{Impulse Response Function with respect to $\sigma^2$: $\sigma^2 = 0.01$ (red line), $\sigma^2 = 0.1$ (green line), and $\sigma^2=0.5$ (cyan line).  }\label{fig:IRFt_t_sigma_different_mu_different}
\end{figure}

\begin{figure}
\centering
\includegraphics[scale = 0.80]{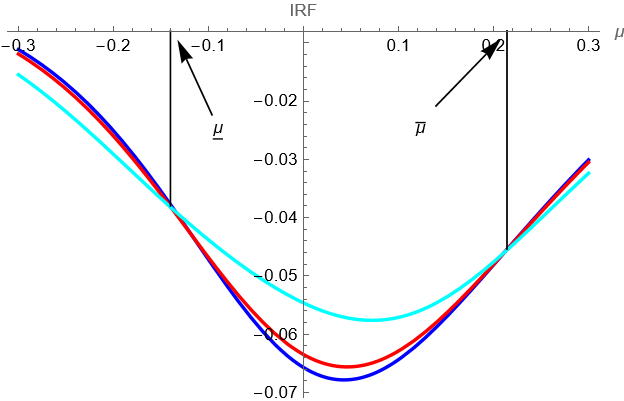}
    \caption{Impulse Response Function at time $1$ with respect to $\mu$ varying $\sigma^2$: $\sigma^2 = 0.01$ (red line), $\sigma^2 = 0.1$ (green line), and $\sigma^2=0.5$ (cyan line).  }\label{fig:IRFt_t_sigma_different_mu_different_regimes}
\end{figure}

{\bf Noise Level $\sigma^2$.} The standard IRF in the VAR framework depends only on the power of matrix $\mathbf{B}$ and on the shock intensity $\boldsymbol{\vec \Delta}$. As mentioned in Section \ref{Methodology}, in the network case the IRF depends also on the noise variance $\sigma^2$ of the $\theta$s. Therefore, we perform a comparative static of the IRF with respect to parameter $\sigma^2$ as $\sigma^2 \in \{0.01,0.1,0.5\}$. 
In the average network (panel (a) of Figure \ref{fig:IRFt_t_sigma_different_mu_different}) we observe that in the short-run, larger $\sigma^2$ leads to smaller effect on the network density. Therefore, the lower the parameter $\sigma^2$, the larger is the time needed to the system to relax back to the equilibrium value. In the cases of dense and sparse network, we observe an opposite behavior since in the short-run the deepest IRF is associated to the highest $\sigma^2$. The main driver of this behavior the value of $\mu$. To show this, we study the IRF at time $t=1$ by varying $\mu$ and choosing three values for $\sigma^2$. The result, displayed in Figure \ref{fig:IRFt_t_sigma_different_mu_different_regimes}, 
shows the existence of two thresholds $\underline{\mu}<0$ and $\bar{\mu}>0$ which define 
sparse and 
dense network cases, respectively. {\color{black}Notice that the thresholds are model parameters' driven: with the parameters of Section \ref{numerical_analysis}, when $\mu \le \underline{\mu}$, the network density is lower that $0.2$ while for $\mu \ge \bar{\mu}$ the network density is higher than $0.9$}. 
Figure \ref{fig:IRFt_t_sigma_different_mu_different_regimes} illustrates that if $\mu \in (\underline{\mu},\bar{\mu})$, the IRF decreases when the parameter $\sigma^2$ increases, whereas the opposite occurs when $\mu \not\in (\underline{\mu},\bar{\mu})$.  We note that the interval $(\underline{\mu},\bar{\mu})$ is not symmetric w.r.t. $\mu =0$ highlighting the asymmetry that characterized the IRF (as seen also in Figure \ref{fig:IRFt_t_Delta_mu_different}) as well as the predominant role of parameter $\mu$ in the shock spreading. Finally, we notice that panel (b) and (c) in Figure \ref{fig:IRFt_t_sigma_different_mu_different} belongs to the extreme cases since the network density is $\delta = 0.95$ (panel (b)) and $\delta = 0.05$ (panel (c)).\\

\begin{figure}
\centering
\includegraphics[scale = 0.80]{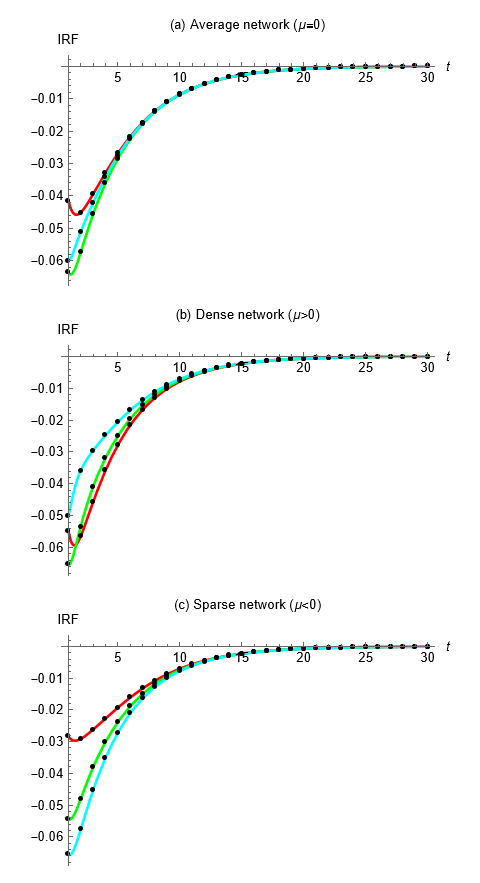}
    \caption{Dependency of IRF on parameter $\mu$ choosing $\theta_0 = -0.75$ (red line), $\theta_0 = 0$ (green line), and $\theta_0=0.75$ (cyan line).}\label{fig:IRFt_t_theta_different_mu_different_regimes}
\end{figure}

{\bf Initial condition $\boldsymbol{\vec \theta_{0}}$.} Finally, we perform a comparative statics of the IRF by assuming that $\boldsymbol{\vec \theta_{0}} \ne \boldsymbol{\vec \theta}_S$.\ As done in the previous analysis, we consider three cases: $\mu \in \{-0.3,0,0.3\}$ so that $\theta_{i,S} \in \{-1.43,0,1.43\}$ $\forall i$ and $\theta_{i,0} \in \{-0.75,0,0.75\}$. Figure \ref{fig:IRFt_t_theta_different_mu_different_regimes} shows that, surprisingly, it is not possible to identify a clear pattern for the dynamics of the IRF. Indeed, the time needed to the system to relax back to the equilibrium depends mainly on (i) the value of the system at the shock time, (ii) the parameter $\mu$, (iii) the sign of the shock. Intuitively, one can argue that the greater the distance between $\theta_{i,0}$ and $\theta_{i,S}$, the more time the system needs to relax back to equilibrium. This is partly true: however, the difference $| \theta_{i,0} - \theta_{i,S}|$ is not the only aspect to be taken into account since, as already widely discussed in the previous sections, the $\Delta$ and $\theta_0$'s signs are essential. Indeed, concordant signs imply a more moderate impact on $\delta$ rather than the case in which $\Delta$ and $\theta_0$ have opposite signs. This last sentence can be observed by looking at cyan line in panel (c) in Figure \ref{fig:IRFt_t_theta_different_mu_different_regimes}.

\section{Model estimation and empirical application to the interbank market}
\label{empirical_application}

In this Section, we first introduce and discuss our novel heuristic estimation procedure of the model described above, showing its effectiveness with numerical simulations, and then we present an empirical illustration to the dynamics of the Italian interbank network e-MID.

\subsection{Estimation Procedure} 
\label{estimation_method}

With the aim of applying the methodology developed in Section \ref{Section_IRF} to a real world application, the first issue relates to the estimation of fitness dynamics. Indeed, while Exponential Random Graphs are extremely flexible and adaptable to a large number of applications, one of the main problems faced by researchers deals with the estimation procedure since the fitness are latent and not observable. Therefore, in this section, we propose a novel estimation procedure for temporal networks induced by latent variables' dynamics. Our approach is based on two steps: first, we estimate the fitness parameters by applying the Maximum Likelihood Estimation (MLE) procedure for each time period. However, despite the MLE has been studied and several algorithms have been implemented, \cite{chatterjee2011} highlights that the procedure does not guarantee consistent estimates in the presence of sparse networks, such as the interbank network (see also \cite{mazzarisi2017}). Additionally, the MLE method produces static estimations, that is, the likelihood function is maximized at each time, without considering the entire dynamics of $\boldsymbol{\vec{\theta}}_t$, i.e.\ regardless the information embedded in past values of $\theta$s. This approach is also known as Single Snapshot Inference (SSI). \\
For these reasons, we propose a novel estimation procedure based on the Kalman Filter (see \cite{khodarahmi2023} for a review) of a state space model composed by an observation and a latent dynamics equation. 
We consider the MLE estimates as noisy observation of the latent parameters which follows the VAR dynamics described above. 

 We remind that the MLE estimations $\boldsymbol{\vec{\Theta}}_t$ obtained by single snapshots and defined as 
\begin{equation}
    \boldsymbol{\vec{\Theta}}_t \in \arg \max P(\mathbf{A}_t|\boldsymbol{\vec \theta}_t)
\end{equation}
are obtained as solution of the system of nonlinear equations
\begin{equation}
    \sum_{i \ne j} \frac{1}{1+e^{-{\Theta}_{i,t} - {\Theta}_{j,t}}} = d_{i,t}, \hspace{3mm} i \in V, \hspace{3mm} t \in \mathcal{T}
\end{equation}
in the undirected case and
\begin{equation}
     \sum_{i < j} \frac{1}{1+e^{-{\Theta}_{i,t}^{in} - {\Theta}_{j,t}^{out}}} = d_{i,t}^{in}, \hspace{3mm} i \in V, \hspace{3mm} t \in \mathcal{T}
\end{equation}
\begin{equation}
     \sum_{i < j} \frac{1}{1+e^{-{\Theta}_{i,t}^{out} - {\Theta}_{j,t}^{in}}} = d_{i,t}^{out}, \hspace{3mm} i \in V, \hspace{3mm} t \in \mathcal{T},
\end{equation}
for the directed case.


 We assume that the observed $\boldsymbol{\vec{{\Theta}}}_t$ are possibly biased noisy estimates of the latent fitnesses $\boldsymbol{\vec{{\theta}}}_t$, i.e. 
$$
\boldsymbol{\vec{{\Theta}}}_t =  \boldsymbol{\vec \gamma} + \mathbf{I} \boldsymbol{\vec{{\theta}}}_t + \boldsymbol{\vec v}_t
$$
 where $\boldsymbol{\vec \gamma}$ is a constant  vector, $\boldsymbol{\vec{{\theta}}}_t$ is the vector collecting the unbiased estimation of fitness vector and $\boldsymbol{\vec v}_t$ is a zero mean Gaussian vector with a constant covariance matrix. The latent state dynamics is described by VAR(1) as in \eqref{Model_VAR_1}.
Therefore, the state space model is described by the following system of equations:
    \begin{equation}
\begin{cases}
     \boldsymbol{\vec{{\theta}}}_t  = &  \boldsymbol{\vec{{\mu}}} + \mathbf{B}  \boldsymbol{\vec{{\theta}}}_{t-1} + \boldsymbol{\vec w}_t  \\
     \boldsymbol{\vec{{\Theta}}}_t = & \boldsymbol{\vec{{\gamma}}} + \mathbf{I}  \boldsymbol{\vec{{\theta}}}_t + \boldsymbol{\vec v}_t 
     \label{model_VAR_estimation}
\end{cases}
\end{equation}
which can be estimated with the Kalman filter. This procedure is termed Kalman-Filter Single Snapshot Inference (KF-SSI). The next Section is devoted to check the effectiveness of our estimation procedure.

\subsection{Numerical Simulation}

We investigate the effectiveness of our estimation procedure presented in Section \ref{estimation_method} by performing numerical simulations of the fitness dynamics in Equation \eqref{Model_VAR_1}. With the purpose of balancing flexibility and computational time, we assume to work in the mean-field framework as in Section \ref{IRF_mean_field}.
We compare the performance of the KF-SSI with the one obtained by performing MLE of the fitnesses, a procedure termed Naive Single Snapshot Inference (N-SSI). Our purpose is to show that KF-SSI achieves superior performances with respect to N-SSI in parameter estimation. \\
We compare the two methods with the following steps:
\begin{enumerate}
    \item Select the set of parameters $\boldsymbol{\vec{\Phi}} = (\mu, a,b,\sigma^2)$.
    \item Simulate $n_{sim}$ time series for $\{\boldsymbol{\vec{\theta}}_t\}_{t \in \mathcal{T}}$ according to dynamics \eqref{Model_VAR_1}.
    \item Construct the network at each time $t$.
    \item Apply the MLE procedure to estimate $\boldsymbol{\vec{{\Theta}}}_t$ at each time $t$. These are the N-SSI estimations.
    \item Estimate the parameters of the KF-SSI models and produce the latent state estimate mean at time $t$ given observations up to and including at time $t$. For the N-SSI we directly fit a VAR(1) model on the MLE fitnesses $\boldsymbol{\vec{{\Theta}}}_t$ to obtain an estimate of the static parameters $\boldsymbol{\vec{\Phi}}$. 
    \item For each estimation method, compute the relative errors in parameters estimation as 
    $$
    RE_{\phi,k} = \frac{\hat{\phi}_{k}-\phi}{\phi}, \hspace{3mm} k \in \{N-SSI,KF-SSI\}, \hspace{2mm} \phi \in \boldsymbol{\vec{\Phi}}
    $$
    where $\hat{\phi}_k$ is the estimation of the parameter $\phi$ in the $k-$th model. We then compute the mean absolute error over the simulations.
    \item Compute the mean absolute error for the estimated fitness dynamics. The mean is obtained by averaging the error over the nodes and the simulations.
 \end{enumerate}

 We simulate $n_{sim}=100$ undirected networks of $n=10$ nodes for $t \in \{1,\dots,100\}$ times from the mean field model with parameters: $a = 0.7$, $b=0.07$, $\sigma = 0.2$, $\mu = -0.07$. 
The mean absolute errors are reported in Table \ref{tab:MAE_estimations}, which confirms that the procedure KF-SSI greatly outperforms N-SSI for all model parameters except for parameter $b$.

\begin{table}[]
    \centering
    \begin{tabular}{rccccc}
      & $\theta_{i,t}$ & $a$ & $b$ & $\mu$ & $\sigma^2$ \\
      \hline \hline
       N-SSI  & 0.57 & 0.605 & 0.009 & 0.311 & 0.375   \\
       KF-SSI &  0.404 & 0.124 & 0.023 & 0.118 & 0.144 \\
    \end{tabular}
    \caption{The mean absolute relative error of {\color{black}the estimates of  fitness dynamics} and of parameters of the network model.}
    \label{tab:MAE_estimations}
\end{table}

\subsection{Data}

In this section, we apply the proposed methodology to the Italian electronic market for interbank deposits (e-MID). The e-MID is an electronic market in the Euro Area where nodes (banks) extend loans to one another for a specified term and/or collateral. The e-MID network has been extensively studied in the last twenty years both in the field of complex networks and in the economics/financial mathematics, see, just to name a few, \cite{iori2006, iori2008,iori2015_emid,cimini2015,mazzarisi2017,barucca2018,mazzarisi2020}. The dataset contains the daily credit transaction between banks, mostly based in Italy, from March 9th, 2012 to February 27th, 2015. Following the approach in \cite{mazzarisi2020}, we aggregate weekly the daily data to create a temporal interbank network. At each time, an arc between two banks $i$ and $j$ indicates that the bank $i$ extend a loan to bank $j$. In doing so, we model the e-MID by the use of a temporal series of unweighted and directed network. 

 As already pointed out by \cite{barucca2018}, the e-MID network is characterized by a sharp decrease in both  number of active banks and traded volume after $2012$ as a result of European sovereign debt crisis and unconventional monetary measures issued by the European Central Bank. To have a stationary sample, we focus our analysis on the period of 40 weeks from January to October, 2014. We select 8 banks  according two criteria: (i) their cumulated in- and out-degree in the time horizon period exceed a fixed threshold set to $100$, (ii) for each time $t$, the induced subgraph\footnote{The induced subgraph of a graph $G=(V,E)$ is another graph formed by a subset of nodes of $G$, $V'$, and all of the edges in $E$ that connect pairs of vertices in $V'$.} does not contain isolated nodes. 


\subsection{Empirical results}

We now consider real data of the e-MID (directed) interbank market by relaxing the mean field assumption
on the structure of matrix $\mathbf{B}$. In doing so, we are able to model the different types of lagged interactions between banks.

\begin{figure}
   \hspace{-3mm}
       \centering
       \includegraphics[scale=0.45]{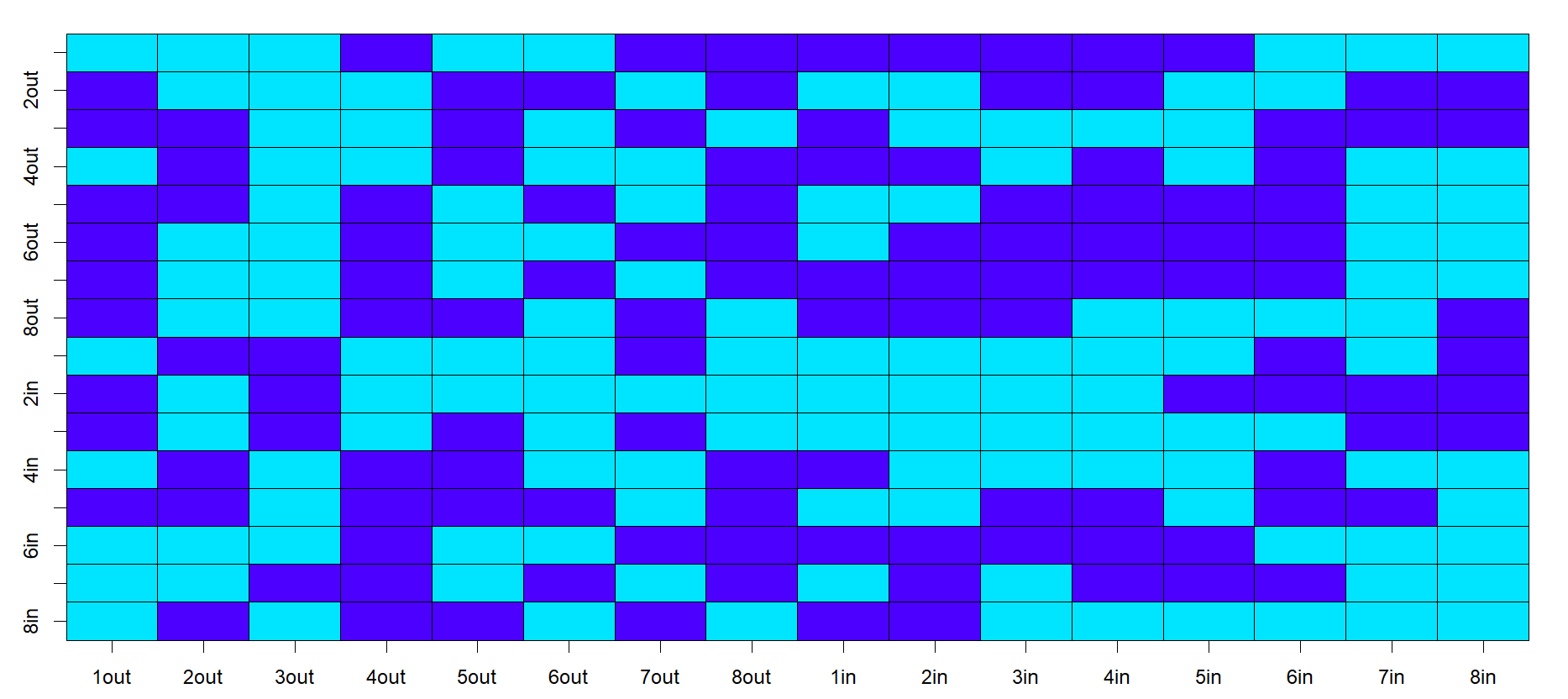}
       \caption{Sign of the entries of the matrix $\mathbf{B}$. Dark blue and light blue entries refer to negative and positive elements, respectively. The $i^{th}$ row/column label shows the $i^{th}$ bank and direction (in or out).}
       \label{fig:plot_sign_B_hat}
   \end{figure}

 Figure \ref{fig:plot_sign_B_hat} reports the sign of the entries of matrix $\mathbf{B}$'s estimation. We observe that a high heterogeneity characterizes the entries of $\mathbf{B}$ both in terms of sign and weight (in absolute value). {\color{black} Indeed, the average entry in matrix $\mathbf{B}$ is $-0.034$ while the standard deviation of matrix $\mathbf{B}$'s entries is $3.06$. Moreover, around half of the entries are positive ($134$ elements over $256$) including the diagonal elements.}
The heterogeneity in the estimated matrix $\mathbf{B}$ highlights three main facts: (i) the relations among banks' connections (degrees) in the analyzed period are diverse, reflecting the heterogeneous behaviour in lending and borrowing interbank loans; (ii) being so different from the matrix of the mean field model, each fitness has a different long-term equilibrium value; (iii) different shocked node may lead to different IRFs. 

 For this specific networks, one can try to give an interpretation to each entry of the matrix $\mathbf{B}$. A positive value of the diagonal elements of $\mathbf{B}$, associated with the couple $(\theta_{i,t}^{in},\theta_{i,t+1}^{in})$ or $(\theta_{i,t}^{out},\theta_{i,t+1}^{out})$, indicates the persistence in the propensity to lend/borrow from the bank $i$, confirming the results, for example, in \cite{mazzarisi2020}, \cite{kobayashi2018} and \cite{freixas2008}). 
A negative value for the matrix element corresponding to the couple $(\theta_{i,t}^{out},\theta_{j,t+1}^{out})$ indicates that a large propensity to lending of bank $i$ at time $t$ leads to a small propensity to lending of bank $j$ the week after. This might be seen as the effect of competition in the credit supply market between banks. Clearly a positive value might be the effect of a global request for credit which is persistent (as seen above) and satisfied by lender $i$ and $j$ at different times. A similar interpretation holds for the matrix element of the couple $(\theta_{i,t}^{in},\theta_{j,t+1}^{in})$ but for the market of credit demand. Finally, a positive value for the matrix element corresponding to the couple $(\theta_{i,t}^{in},\theta_{j,t+1}^{out})$ indicates that a large propensity to borrowing of bank $i$ at time $t$ is associated with a large propensity to lending of bank $j$ in the next period. In part this might be due to the fact that $\theta_{i,t}^{in}$ is persistent, thus also $\theta_{i,t+1}^{in}$ will be likely large, creating opportunity for lending of bank $j$. A similar interpretation, but with opposite sign could be given to the value of the matrix element corresponding to $(\theta_{i,t}^{out},\theta_{j,t+1}^{in})$. Although these interpretations are tentative and should be validated with more structural models, their meaning in terms of lagged correlation between the propensity to lending/borrowing can be certainly given as they follow from the significance of the VAR coefficient.

\begin{figure}
    \centering
    \includegraphics[scale = 0.85]{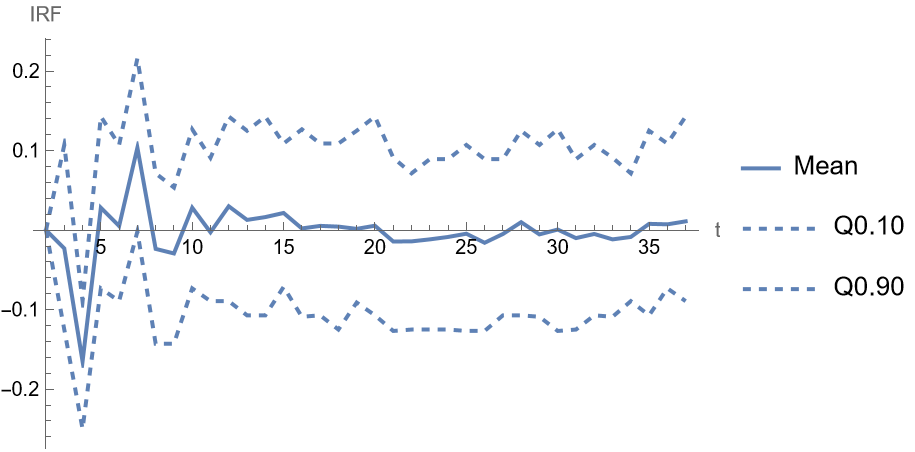}
    \caption{Impulse Response Function for e-MID network with negative $\Delta = -0.3$. Solid line represents the simulations' mean, dotted lines represent the $10^{th}$ and $90^{th}$ percentiles, respectively.}
    \label{fig:emid_shock_network}
\end{figure}


Based on these observations, we computed the IRF by shocking the node with the largest out-fitness at time $\tau$ and to compute the  IRF conditional on different information sets. Specifically, we simulate $500$ IRFs for the e-MID interbank network conditional to shock $\Delta  = -0.3$ on the out- fitness of the node which presents the maximum out-degree in the week from 02/14/2014 to 02/21/2014 (fifth week). From economically point of view, this scenario considers the case when the larges lender bank in the market significantly reduces the amount lent to the other banks. The mean IRF and the $10^{th}$ and $90^{th}$ percentiles across simulations are plotted in Figure \ref{fig:emid_shock_network}. To compare the simulations results, we assume that the fitnesses values at the shock time are provided by KF-SSI model specification in \eqref{model_VAR_estimation}. \\
Interestingly, the largest negative variation occurs at a later time than $\tau +1$, {\color{black}as already observed in the first comparative statics of Section \ref{numerical_analysis}}. This {\color{black}suggests} that there exists a lagged effect in the shock spreading in the interbank market. Subsequently, the IRF presents a significant positive change which leads the IRF to be positive. Finally, the shock is completely absorbed and the network density relaxes back to the pre-shock value. 

Finally, we study how the IRF in the network case depends on the network state at $\tau$, i.e.\ on $\boldsymbol{\vec \theta}_{\tau}$ and on the shocked node. To this end, we consider different initial conditions {\color{black}as done also in Figure \ref{fig:IRFt_t_theta_different_mu_different_regimes} in Section \ref{numerical_analysis}}. 
We observe that in these cases, {\color{black}as already observed in Figure \ref{fig:IRFt_t_theta_different_mu_different_regimes},} the IRF {\color{black}patterns} are close to the one shown in Figure \ref{fig:emid_shock_network}.  {\color{black}This result depends on three main factors: (i) on the VAR dynamics that we impose to $\boldsymbol{\vec \theta}$'s in Eq.\ \eqref{Model_VAR_1}, (ii) the estimated fitness unconditional mean is negative, leading the network to be sparse and (iii) the estimated volatility is low. Nevertheless, we observe a variation in the IRF depending on the shocked node: the higher the fitness value of the shocked node at the shock time, the lower the effect on the network density.} \\
The presented approach has potential interest for policy makers and supervising authorities. Indeed, the results suggest that the dynamics of the IRF is mainly determined by the matrix $\mathbf{B}$ of the VAR model. 
The estimation of such a matrix allows policymakers to assess the impacts of both systemic exogenous shocks and different economic/financial policies. Especially in the first case, knowing the entries of matrix $\mathbf{B}$ allows the policymaker to construct dynamic scenarios of the evolution of the interbank network under different stress conditions and possibly to  minimize the effect of systemic risk. 

\section{Conclusion}
In this paper, we presented a dynamical model for temporal networks and we introduced the first Impulse Response Function for networks metric. The aim of the model and of the analysis is to assess how an external shock propagates in a temporal network as well as to establish its resilience. This is done by extending the celebrated configuration or fitness model for static network to a fully dynamical setting. The model associates each node to a pair of latent variables that represents its propensity to create in- and out-links in the network. We assume that these variables evolves stochastically in time according to a VAR(1) process to model the lead-lag relation between the linking propensities of the nodes. The consequence is that the network and all its metrics are stochastic. In such a framework, we analyze the variation of a given network metric after a shock in the fitness of a node. Additionally, we analyze how the system relaxes back to the equilibrium state examining also the recovery time. We provide a general formulation for the IRF for a general network metric. We focus on the network density and we determine a closed-form solution for its IRF under the assumption of fitnesses' homogeneity. To assess the role of each model's parameter, we make a comparative statics considering three different values for the network density corresponding to average, dense, and sparse networks. We propose an econometric approach to estimate the fitnesses' dynamics based on the Kalman filter. Finally, we apply our methodology to the e-MID network. The results highlight not only the different dynamics that characterizes the lending/borrowing mechanism in the e-MID network but also specify the role of the banks in the shock spreading as well as the pattern through which a shock on financial network is absorbed. 

The presented framework can be extended in different directions. First, other shock scenarios could be considered. Although here we focused on a shock hitting a single node, one can easily extend our computations to the case when a subset or even all the nodes are simultaneously affected by a global shock. This would allow to study the shock propagation and resilience of a network when an exogenous shock hits part of all the system. Second, other network metrics can be investigated to understand the effect of shocks on their temporal evolution. Proposition 1 gives a general answer to the problem, but the detailed computations must be spelled out. Finally, one it might be interesting to develop alternative and more rigorous estimation techniques able to infer the fitness dynamics in sparse networks. We leave these interesting questions for future research.

\section*{Acknowledgement}
The authors gratefully acknowledge financial support from "SoBigData.it", which receives funding from European Union – NextGenerationEU – PNRR – Project: “SoBigData.it – Strengthening the Italian
RI for Social Mining and Big Data Analytics” – Prot. IR0000013 – Avviso n. 3264 del 28/12/2021, and
"NetRes - Network analysis of economic and financial resilience" (PRO3 Scuole). 
The author acknowledge partial support by the European Program scheme ‘INFRAIA-01-2018-2019: Research and Innovation action’, grant agreement n. 871042 ’SoBigData++: European Integrated Infrastructure for Social Mining and Big Data Analytics’.

\bibliography{Myref}

\appendix 

\section{Derivation of network density under Gaussian distributed fitnesses}
\label{Appendix_A}
In this appendix, we show the passages to derive Formula \eqref{density_media_general}. The case with different means is based on the same procedure. As stated in Section \ref{Methodology}, we assume that the vector $\boldsymbol{\vec \theta} $ is a multivariate gaussian distribution which entries have all the same mean $m$, variance $s^2$ and correlation factor $r$. We prove that the expected density of a network following a fitness model is 
\begin{eqnarray}
\mathbb{E[\delta ]} &\mathbb{=}&\frac{1}{2s^{2}\pi \sqrt{1-r ^{2}}}%
\int_{-\infty}^{\infty} \int_{-\infty}^{\infty} \frac{1}{1+e^{-\left( \theta_i +\theta_{j}\right) }}e^{-\frac{1}{2s^{2}\left(
1-r ^{2}\right) }\left( \left( \theta_{i}-m \right) ^{2}-2r \left( \theta_i -m
\right) \left( \theta_{j}-m \right) +\left( \theta_{j}-m \right) ^{2}\right) }d\theta_i d\theta_j \label{integral_appendix_A} \\
& = & I(2m, 2\sigma^2(1+r)) \notag
\end{eqnarray}
where $I(2m, 2\sigma^2(1+r))$ is the 
normal-logistic integral 
 We first rearrange the numerator of the exponent inside the integral
signs as
\begin{eqnarray}
&&\left( \theta_{i}-m \right) ^{2}-2r \left( \theta_{i}-m \right) \left( \theta_{j}-m \right)
+\left( \theta_{j}-m \right) ^{2} \notag
\\ &&=   2m ^{2}-2\theta_{j}m -2\theta_{i}m -2m
^{2}r +\theta_{i}^{2}+\theta_{j}^{2}-2\theta_{i}\theta_{j}r +2\theta_{i}\allowbreak m r +2\theta_{j}m r \notag \\ 
&&=\theta_{i}^{2}+\theta_{j}^{2}-2m \left( \theta_{i}+\theta_{j}\right) +2m r \left( \theta_{i}+\theta_{j}\right) +r
\left( \frac{-\left( \theta_{i}+\theta_{j}\right) ^{2}}{2}+\frac{\left( \theta_{i}-\theta_{j}\right) ^{2}}{2}%
\right) +2m ^{2}\left( 1-r \right) \notag \\ 
&&=\frac{\left( \theta_{i}+\theta_{j}\right) ^{2}+\left( \theta_{i}-\theta_{j}\right) ^{2}}{2}-2m \left(
\theta_{i}+\theta_{j}\right) \left( 1-r \right) +r \left( \frac{-\left( \theta_{i}+\theta_{j}\right) ^{2}}{%
2}+\frac{\left( \theta_{i}-\theta_{j}\right) ^{2}}{2}\right) +2m ^{2}\left( 1-r \right) \label{numeratore_finale}
\end{eqnarray}
where in the last passage, we use the fact that 
$\theta_{i}^{2}+\theta_{j}^{2}=\frac{\left( \theta_{i}+\theta_{j}\right)^{2}+\left( \theta_{i}-\theta_{j}\right) ^{2}}{2}$ and $-2\theta_{i}\theta_{j}=\frac{-\left( \theta_{i}+\theta_{j}\right) ^{2}}{2}+\frac{\left( \theta_{i}-\theta_{j}\right) ^{2}}{2}$. \\

Then we use a change of variables $z=\theta_{i}-\theta_{j}$ e $w=\theta_i +\theta_{j}$ so that Eq.\ \eqref{numeratore_finale} can be written as
\begin{eqnarray*}
&&\frac{w^{2}+z^{2}}{2}-2m w\left( 1-r \right) +r \left( -\frac{w^{2}%
}{2}+\frac{z^{2}}{2}\right) +2m ^{2}\left( 1-r \right) + + \frac{1}{2} z^{2}\left(
1+r \right) \\
&=& \frac{1}{2} \left( w-2m \right) ^{2}\left( 1-r \right) + \frac{1}{2} z^{2}\left(
1+r \right),
\end{eqnarray*}
and the double integral in Eq.\ \eqref{integral_appendix_A} can be written in terms of $w$ and $z$ as
$$
\frac{1}{2}\frac{1}{2s^{2}\pi \sqrt{1-r ^{2}}}\int_{-\infty}^{\infty} \int_{-\infty}^{\infty} \frac{1}{%
1+e^{-w}}e^{-\frac{1}{4s^{2}\left( 1-r ^{2}\right) }\left( \left(
w-2m \right) ^{2}\left( 1-r \right) +z^{2}\left( 1+r \right) \right)
}dwdz
$$
where the term $\frac{1}{2}$ is the Jacobian of the variables change $\theta_{i}=\frac{w+z}{2}$ and $\theta_{j}=\frac{w-z}{2}$.\\

Finally we observe that the double integral can be written as the product of two integrals, one of whom is the normal-logistic integral. Therefore, the expected density of the network can be expressed as 
\begin{eqnarray*}
&&\frac{1}{2}\frac{1}{2s^{2}\pi \sqrt{1-r ^{2}}}\int_{-\infty}^{\infty} \int_{-\infty}^{\infty} \frac{1}{%
1+e^{-w}}e^{-\frac{1}{4s^{2}\left( 1-r ^{2}\right) }\left( \left(
w-2m \right) ^{2}\left( 1-r \right) +z^{2}\left( 1+r \right) \right)
}dwdz \\
&& \\
&=&\frac{1}{2}\frac{1}{2s^{2}\pi \sqrt{1-r ^{2}}}\int_{-\infty}^{\infty} e^{-\frac{%
z^{2}\left( 1+r \right) }{4s^{2}\left( 1-r ^{2}\right) }}dz \int_{-\infty}^{\infty}
\frac{1}{1+e^{-w}}e^{-\frac{\left( w-2m \right) ^{2}\left( 1-r \right) 
}{4s^{2}\left( 1-r ^{2}\right) }}dw \\
&& \\
&=&\frac{1}{2}\frac{\sqrt{2}\sqrt{\pi }s\sqrt{1-r }}{2\sigma
^{2}\pi \sqrt{1-r ^{2}}}\int_{-\infty}^{\infty} \frac{1}{1+e^{-w}}e^{-\frac{\left( w-2m \right) ^{2}}{4s^{2}\left( 1+r \right) }}dw \\
&& \\
&=&\frac{1}{2}\frac{\sqrt{1-r }}{s\sqrt{2}\sqrt{\pi }\sqrt{1-r
^{2}}}\int_{-\infty}^{\infty} \frac{1}{1+e^{-w}}e^{-\frac{\left( w-2m \right) ^{2}}{4\sigma
^{2}\left( 1+r \right) }}dw \\
&& \\
&=&I\left( 2m ,2s^{2}\left( 1+r \right) \right) 
\end{eqnarray*}

\section{Proof of Proposition 1}
\label{Appendix_B_new}
In this appendix, we prove the Proposition \ref{proposition1} in Section \ref{Section_IRF}. 
Let $f(\mathbf{A})$ be a network metric and indicating with $\{\boldsymbol{\vec{\theta}}\}_{0:\tau}$ the past history of the fitness vector, it is possible to write explicitly the expected value of the network metric $f(\mathbf{A}_t)$ conditioned to the history of $\boldsymbol{\vec \theta}_{t}$ from $0$ to time $\tau$ as
\begin{equation}
{\mathbb E}[f(\mathbf{A}_t)|\{\boldsymbol{\vec{\theta}}\}_{0:\tau}]=
\int {\mathbb E}\left[f(\mathbf{A}_t)|\boldsymbol{\vec{\theta}}_t\right] P\left(\boldsymbol{\vec{\theta}}_t|\{\boldsymbol{\vec{\theta}}\}_{0:\tau} \right) d\boldsymbol{\vec{\theta}}_t
\label{conditioned_expected_value}
\end{equation}
Given the Markovianity and Gaussianity of our setting, Eq.  \eqref{conditioned_expected_value} becomes
\begin{eqnarray}
{\mathbb E}[f(\mathbf{A}_t)|\{\boldsymbol{\vec{\theta}}\}_{0:\tau}]=
\int {\mathbb E}\left[f(\mathbf{A}_t)|\boldsymbol{\vec{\theta}}_t\right] P\left(\boldsymbol{\vec{\theta}}_t|\boldsymbol{\vec{\theta}}_{\tau} \right) d\boldsymbol{\vec{\theta}}_t \nonumber\\
=\int {\mathbb E}\left[f(\mathbf{A}_t)|\boldsymbol{\vec{\theta}}_t\right] {\mathcal N}\left( \boldsymbol{\vec{\theta}}_t;\boldsymbol{\vec \mu}_t,\boldsymbol{\Sigma}_t\right)d\boldsymbol{\vec{\theta}}_t
\label{simplified_cond_expe_value}
\end{eqnarray}
where $\boldsymbol{\vec\mu}_t$ and $\boldsymbol{\Sigma}_t$ are, respectively, the conditional mean and variance of $\boldsymbol{\vec \theta}_t|\{\boldsymbol{\vec{\theta}}\}_{0:\tau}$  in Eq.s \eqref{eq:mut} and \eqref{eq:sigmat}. 
By Equation \eqref{IRF_general}, the IRF for the network metric $f(\mathbf{A}_t)$ is defined as 
$$
{\boldsymbol I}{\boldsymbol R}{\boldsymbol F}^f(t;\boldsymbol{\vec{\Delta \theta}})={\mathbb E}[f(\mathbf{A}_{t+\tau})| \boldsymbol{\vec{\theta}}_{\tau}+\boldsymbol{\vec{\Delta \theta}},\boldsymbol{\vec{\theta}}_{\tau-1}... ]-{\mathbb E}[f(\mathbf{A}_{t+\tau})| \boldsymbol{\vec{\theta}}_{\tau},\boldsymbol{\vec{\theta}}_{\tau-1}... ]
$$
and by using \eqref{simplified_cond_expe_value}, it is possible to write the IRF as a Gaussian integration as 
$$
{\boldsymbol I}{\boldsymbol R}{\boldsymbol F}^f(t;\boldsymbol{\vec{\Delta \theta}})=
\int {\mathbb E}\left[f(\mathbf{A}_t)|\boldsymbol{\vec{\theta}}_t\right] 
\cdot \left[ {\mathcal N}\left( \boldsymbol{\vec{\theta}}_t;\boldsymbol{\vec \mu}_t+\mathbf{B}^{t-\tau}\boldsymbol{\vec{\Delta \theta}},\boldsymbol{\Sigma}_t\right)-{\mathcal N}\left( \boldsymbol{\vec{\theta}}_t;\boldsymbol{\vec \mu}_t,\boldsymbol{\Sigma}_t\right)\right]d\boldsymbol{\vec{\theta}}_t.
$$

\section{Power of VAR matrix in mean-field model}
\label{Sec_power_matrix_new}

As assumed in Section \ref{numerical_analysis}, we deal with a full matrix $\mathbf{B}$ whose diagonal elements are all equal to $a$ and the off-diagonal elements are equal to $b$. Observing that such matrix can be written as $\mathbf{B}= (a-b) \mathbf{I} + b \mathbf{1}$, where $\mathbf{I}$ is the identity matrix and $\mathbf{1}$ is the matrix which elements are all one, then the $t-$power of such matrix can be written as
\begin{equation}
    \mathbf{B}^t = \mathbf{I}(a-b)^t + \frac{\lambda_1^t - (a-b)^t}{n} \mathbf{1}.
    \label{power_mat_B}
\end{equation}
We use Eq. \eqref{power_mat_B} to write analytically the conditional mean and the conditional variance in Equations \eqref{eq:mut} and \eqref{eq:sigmat}, respectively. For simplicity, we assume that $\tau =0$ and first, we explicitly write the following geometric sum
$$
\mathbf{I} + \mathbf{B}^1 + \mathbf{B}^2 + \cdots + \mathbf{B}^{t-1} = \sum_{k=0}^{t-1}\mathbf{B}^k.
$$
which is equal to
$$
\mathbf{I} \frac{1-(a-b)^t}{1-(a-b)} +  \mathbf{1} \left[ \frac{1-\lambda_1^t}{1-\lambda_1} - \frac{1-(a-b)^t}{1-(a-b)} \right]\frac{1}{n}.
$$
By same reasoning, it is possible to derive the power function 
$$
\mathbf{I} + \mathbf{B}^2 + \mathbf{B}^4 + \cdots + \mathbf{B}^{2(t-1)} = \sum_{k=0}^{t-1}\mathbf{B}^2k = 
\mathbf{I} \frac{1-(a-b)^{2t}}{1-(a-b)^2} +  \mathbf{1} \left[ \frac{1-\lambda_1^{2t}}{1-\lambda_1^2} - \frac{1-(a-b)^{2t}}{1-(a-b)^2} \right]\frac{1}{n}.
$$

\end{document}